\let\latexaddtocontents\addtocontents
\documentclass[a4paper,twocolumn,11pt,unpublished]{quantumarticle}
\let\addtocontents\latexaddtocontents

\pdfoutput=1
\usepackage[utf8]{inputenc}
\usepackage[english]{babel}
\usepackage[T1]{fontenc} 
\usepackage{braket}
\usepackage{amssymb}
\usepackage{amsmath}
\usepackage{mathtools}
\usepackage{hyperref}

\usepackage{tikz}
\usepackage{pgfplots}
\pgfplotsset{compat=newest}
\usetikzlibrary{calc}
\usepackage{xfrac}

\DeclareMathOperator{\tr}{\text{Tr}}

\newcommand{\beginappendix}[1]{%
        \setcounter{equation}{0}
        \renewcommand{\theequation}{#1\arabic{equation}}%
     }

\usepackage[
    backend=biber,
    citestyle=numeric,
    bibstyle=numeric,
    url=false,
    maxbibnames=6,
    sorting=none
]{biblatex}
\renewbibmacro{in:}{%
  \ifentrytype{article}{}{\printtext{\bibstring{in}\intitlepunct}}}
\DeclareFieldFormat[article]{title}{#1}
\DeclareFieldFormat[article]{volume}{\bf{#1}}
\begin{document}
\title{High-dimensional entanglement certification: bounding relative entropy of entanglement in $2d+1$ experiment-friendly measurements}

\author{Alexandria J. Moore}
\affiliation{School of Electrical and Computer Engineering and Purdue Quantum Science and Engineering Institute, Purdue University, West Lafayette, IN 47907, United States of America}
\email{moore428@purdue.edu}
\orcid{0000-0001-7385-4435}
\author{Andrew M. Weiner}
\affiliation{School of Electrical and Computer Engineering and Purdue Quantum Science and Engineering Institute, Purdue University, West Lafayette, IN 47907, United States of America}
\orcid{0000-0002-1334-8183}
\maketitle

\begin{abstract}
	Entanglement -- the coherent correlations between parties in a joint quantum system -- is well-understood and quantifiable in the two-dimensional, two-party case. Higher ($>$2)-dimensional entangled systems hold promise in extending the capabilities of various quantum information applications. Despite the utility of such systems, methods for quantifying high-dimensional entanglement are more limited and experimentally challenging. We review entanglement certification approaches and the large number of -- often difficult -- measurements required to apply them.  We present a novel certification method whose measurement requirements scale linearly with subsystem dimension (scaling with the square-root of the system dimension) and which requires only a single complex measurement. The certification method places a lower-bound on the relative entropy of entanglement of any maximally correlated state thereby certifying system entanglement. A lower bound is also shown for any maximally correlated state in the presence of noise -- the expected experimental case. We discuss experimental realization of all required measurements.
\end{abstract}

Sufficiently large quantum computers are capable of solving problems intractable for their classical counterparts, motivating extensive work towards their realization. Entanglement is a fundamental resource used by quantum computers and quantum communications. Accordingly, the generation and certification of entangled states is key to various quantum technologies, such as quantum key distribution and quantum teleportation~\cite{nielsen2002quantum}. Further, high-dimensional entanglement is valuable for a number of quantum information applications, including superdense coding~\cite{sazim2013study_superdense}, qudit teleportation~\cite{Zhang2019_quditTeleportation}, and some quantum key distribution protocols~\cite{cerf2002security_hdQKD, mirhosseini2015high_hdQKD, mafu2013higher_hdQKD}. However, certifying entanglement in known high-dimensional mixed quantum states can prove challenging~\cite{Huber2018_CertificationReview}. For unknown high-dimensional states, certification complexity -- if possible -- increases rapidly as a large number of difficult measurements are required.

For a special class of high-dimensional states -- known as Schmidt correlated or maximally correlated states -- and noisy versions thereof, we present an experimentally-friendly set of measurements for certifying entanglement. The method lower-bounds the relative entropy of entanglement of a $d^2$ high-dimensional bipartite quantum state (i.e. a system comprised of two $d$-dimensional parties) using only $2d + 1$ measurements: $2d$ simple measurements in the standard basis and a single demanding measurement -- the state's purity. We derive these lower bounds and outline a measurement strategy.

Section~\ref{sec:background} examines the measurements necessary to certify entanglement through existing methods, reviews key proprieties of quantum systems, and defines the class of high-dimensional states of interest: maximally correlated states. Section~\ref{sec:ree} defines relative entropy of entanglement -- the entanglement monotone applied here -- and Section~\ref{sec:Dree_max} derives lower bounds for this quantity. Section~\ref{sec:D_noise} considers noisy maximally correlated states and lower bounds their entanglement using the relative entropy of entanglement. Section~\ref{sec:ExperimentalApplication} outlines the required measurements in the noise-free and noisy cases and shows how to make a direct purity measurement. Section~\ref{sec:discussion} considers the dimensionality of the system's entanglement. In addition, it considers the relative entropy of entanglement's (and the underlying von Neumann entropy's) relation to other quantifiers of entanglement.

As quantum communication tasks which demand spatially disparate high-dimensional entanglement carriers are best served by entangled photons~\cite{guo2019advances}, some discussion around the motivation and application of our method will be presented in terms of photonic systems; however, the method is applicable to any bipartite high-dimensional system.

\section{Background and motivation}\label{sec:background}
\subsection{Entanglement certification}
(Nomenclature note: for a quantum state $\rho$, we will call the relatively simple measurements of $\braket{i,j | \rho | i,j}\forall i, j$ correlation measurements and the more challenging $\left\lvert \braket{i,j | \rho | k,l}\right\rvert$ for $i\neq k$ or $j \neq l$ interferometric measurements.)

Entanglement quantification and certification is well-understood for pure, 2-dimensional, bipartite states. When considering mixed, high-dimensional systems quantification becomes vastly more complex and experimentally challenging~\cite{Huber2018_CertificationReview, PhysRevA.65.032314}.  Some well-known quantifiers high-dimensional entanglement include entanglement of formation~\cite{PhysRevA.54.3824}, Schmidt number~\cite{Terhal2000_SchmidtNumberDef}, mutual information~\cite{Huang2016_MutualInformationCertification}, G-concurrence~\cite{Siewert2019_gconcurrence}, and (logarithmic) negativity~\cite{PhysRevA.58.883, PhysRevLett.95.090503}.

Using these quantifiers to certify entanglement poses challenges. Several approaches may begin with a full quantum state tomography on the quantum state $\rho$ of interest, but the number of measurements required scales unfavorably with the dimension of $\rho$~\cite{Kwiat2005_QST} and further may not be feasible experimentally. As full state $\rho$ knowledge is inaccessible, entanglement certification typically relies on a more limited set of measurements. Take for example G-concurrence which -- for high-dimensional bipartite states -- also requires measurements scaling with $d^2$, half of which are interferometric. Important high-dimensional bipartite systems include multi-photon systems entangled by frequency degree of freedom~\cite{kues2019quantum}. Various other schemes which have been devised to certify the entanglement of formation~\cite{Huber, Wu}, Schmidt number, or related quantities will similarly require a large number of interferometric measurements.

By contrast, lower-bounding mutual information has been accomplished by measuring a system in two mutually unbiased bases~\cite{Huang2016_MutualInformationCertification, HuberErker2017_MutualBasisCertificaiton}, necessitating $d$ correlation measurements in each choice of basis. The first basis may be the computational basis, but the second would then be a highly transformed version of the first. For systems where this transformation is nontrivial, which include photonic high-dimensional frequency- and time-entangled states, we consider the $d$ measurements performed in the mutually unbiased basis to be interferometric. Methods which require measurements made in random bases~\cite{Zoller2018_RandomQuenches, Zoller2020_QDeviceVar} are similarly challenging.

Entangled systems with limited ability to transform into arbitrary bases and for which interferometic measurements are similarly restricted are therefore drastically limited in their certifiable entanglement under standard methods. For this reason, the certification method presented here has been designed to work with only $2d$ correlation measurements and a single more challenging measurement: purity, whose direct measurement is discussed in Section~\ref{sec:Purity}.

\subsection{Purity, separability, and maximally correlated states}\label{sec:Schur}
Entangled states are defined as states which are not separable. A pure, bipartite state $\ket{\psi}$ with subsystems $A$ and $B$ is considered separable if (and only if) it can be written
\begin{align}\label{eq:1}
	\ket{\psi} = \ket{\psi}_A \otimes \ket{\psi}_B.
\end{align}
A mixed quantum state represented by density operator
\begin{align}\label{eq:basic}
	\rho = \sum_{m} w_m \ket{\psi_m}\bra{\psi_m} \text{where } w_m \in \mathbb{R}_{>0}
\end{align}
is considered separable if and only if it can be written as a weighted sum of separable pure states (meaning all $\ket{\psi_m}$ take the form of Equation~\ref{eq:1}) with positive, real weights $w_m$. Separable states lack the coherence which allows entangled states their quantum advantage. Note that all valid quantum states -- separable or not -- take the form of Equation~\ref{eq:basic} with $\sum_m w_m = 1$, $\braket{\psi_m | \psi_m} = 1\forall m$ and are consequently positive semidefinite Hermitian matrices with trace one.

Both entangled and separable states may be correlated in some basis. A bipartite state is correlated when measurement of subsystem $A$ provides additional information on the state of subsystem $B$ (or vice versa). I.e. the measurement results of systems $A$ and $B$ are not independent. However, a key property of highly-entangled states are their strong correlations in multiple bases. A separable system which is highly correlated in one basis will be weakly or completed uncorrelated in a mutually unbiased basis, whereas highly entangled states can be maximally correlated in two (or more) mutually unbiased bases. Entangled states achieve this duality by having high coherence while maximally correlated separable states are entirely incoherent. A state $\rho$'s coherence can be quantified by its purity~\cite{Jaeger_purity},
\begin{align}\label{eq:purity}
	\mathcal{P}(\rho) \equiv \text{Tr}(\rho^2)
\end{align}
where $\rho$ is a density operator (Hermitian, trace-1) representing a quantum state and $\tr (\cdot)$ is the trace operator which is the sum of the diagonal elements of its argument. Separable states may be either highly pure or highly correlated, but not both. Whereas entangled states are characterized by their simultaneous strong correlations in multiple bases~\cite{Huang2016_MutualInformationCertification} and consequently high purity. More succinctly: a highly-correlated state with low purity indicates separability; a highly-correlated state with high purity indicates entanglement.

We exploit the unique relationship between correlations and purity enjoyed by highly-entangled states to lower bound system entanglement in the case of maximally correlated states, defined presently: Let $\rho$ be a density matrix of a bipartite system, where each subsystem is $d$-dimensional. The system is called maximally correlated~\cite{Hiroshima2004_SchmidtCorrelatedDefinition} if and only if
\begin{align}\label{eq:MCDefinition}
	\begin{split}
	\big(\bra{i}_A \otimes \bra{j}_B\big) \rho \big(\ket{i}_A \otimes \ket{j}_B \big) = \braket{i,j|\rho | i,j}\\
	 = 0 \forall i\neq j.
	 \end{split}
\end{align}
Equivalence between Equation~\ref{eq:MCDefinition} and maximally correlated states as defined by~\cite{Hiroshima2004_SchmidtCorrelatedDefinition} is shown in Appendix~\ref{appendix:mc}. The only (potentially) nonzero diagonal entries of a maximally correlated system $\rho_{MC}$ will be denoted
\begin{align}\label{eq:zetaDef}
	\zeta_k \equiv \braket{k,k | \rho_{MC} | k,k}.
\end{align}
The array of all $d$ of these diagonal values will be written
\begin{align}\label{eq:zetaDef2}
	\vec{\zeta} = \braket{\zeta_1, \zeta_2, ..., \zeta_d}
\end{align}
and similar notation will be used for other arrays of values (e.g. eigenvalues).

As all valid density operators are trace-1, keep in mind
\begin{align}
	\sum_{k=1}^d \zeta_k = 1.
\end{align}


All quantum states can be written as a weighted sum of the outer product of pure states, as shown in Equation~\ref{eq:basic}. Such a sum is known as a decomposition of $\rho$ and is not unique. All $\rho$ will have a particular decomposition, known as an eigendecomposition, such that
\begin{align}\label{eq:eigendecomp}
	\rho = \sum_k \phi_k \ket{\phi_k}\bra{\phi_k}
\end{align}
where 
	$\rho \ket{\phi_k} = \phi_k \ket{\phi_k}$.
That is, $\rho$ can be written as an eigenvalue-weighted sum of its (orthonormal) eigenvectors:
\begin{align}\label{eq:eigenterm2}
	\braket{\phi_n | \phi_m} = \delta_{n,m}
\end{align}
and its nonnegative, real eigenvalues, which sum to one,
\begin{align}\label{eq:eigcons}
	\sum_k \phi_k = 1.
\end{align}
Applying this decomposition, we recognize that the purity of $\rho$ (Equation~\ref{eq:purity}) is the squared sum of its eigenvalues:
\begin{align} \label{eq:puritycons}
	\mathcal{P}(\rho) &\equiv \tr(\rho^2)
			= 	\tr (\sum_n \sum_m \phi_n \phi_m \ket{\phi_n}\braket{\phi_n | \phi_m} \bra{\phi_m}) \nonumber   \\
			&=	\tr(\sum_n \phi_n^2 \ket{\phi_n}\bra{\phi_n}) \nonumber \\
			&=	\sum_k \bra{\phi_k }\big(\sum_n  \phi_n^2 \ket{\phi_n}\bra{\phi_n} \big) \ket{\phi_k}\nonumber \\
	\mathcal{P}(\rho)
			&=	\sum_k \phi_k^2.
\end{align}

By the Schur-Horn theorem~\cite{10.2307/2372705} for any Hermitian matrix $\rho$ (which all valid density operators are), the eigenvalues of $\rho$ will always majorize~\cite{peajcariaac1992convex} the diagonal values of $\rho$. That is, for a $d^2$ system $\rho$, if $\vec{\phi}$ are its $d^2$ eigenvalues and $\vec{\iota}$ are its $d^2$ diagonal entries (of the form $\iota_{d(i-1) + j} = \braket{i,j |\rho |i,j}$), then $\vec{\phi}$ will majorize $\vec{\iota}$ ($\vec{ \phi } \succ \vec{ \iota }$), meaning
\begin{align}\label{eq:majone}
	\begin{split}
	\sum_{k=1}^m \phi_{[k]} \geq \sum_{k=1}^m \iota_{[k]},
	\text{ for } m=1, 2, 3, ... 
	\end{split}
\end{align}
with equality for $m=d^2$. Where the bracket $_{[\cdot]}$ notation indicates the elements have been re-indexed in non-increasing order. I.e. it is an indexing of array $\vec{\phi}$ such that $\phi_{[1]} \geq \phi_{[2]} \geq ... \geq \phi_{[d^2]}$ and likewise for $\vec{\iota}$. In addition, the bracket notation will define $\phi_{[>d^2]}=0$. This is useful in the general case when $\vec{\phi}$ and $\vec{\iota}$ are different lengths (e.g. length $M$ and length $N$ where $M\neq N$), in which case majorization requires that Equation~\ref{eq:majone} is an equality for $m=\max(N, M)$.

As a maximally correlated $\rho_{MC}$ will have only $d$ nonzero diagonal entries $\vec{\zeta}$ (Equation~\ref{eq:zetaDef2}), Equation~\ref{eq:majone} becomes
\begin{align}\label{eq:majcons}
	\begin{split}
	\sum_{k=1}^m \phi_{[k]} \geq \sum_{k=1}^m \zeta_{[k]}, \text{ for } m=1, 2, 3, ....
	\end{split}
\end{align}
with equality for $m=\max(d, d^2)= d^2$. As $\vec{\zeta}$ is length-$d$ and as all eigenvalues $\vec{\phi}$ are nonnegative, equality for $m=d^2$ implies equality for $m\geq d$, meaning the array of eigenvalues $\vec{\phi}$ has $d$ or fewer nonzero values for any maximally correlated $\rho_{MC}$. Redefining $\vec{\phi}$ as the array of nonzero eigenvalues (rather than as the array of all eigenvalues), the length $\lvert \vec{\phi}\rvert \leq d$. Any state with $K$ or fewer nonzero eigenvalues $\rho_{\lvert\vec{\phi}\rvert \leq K}$ will have 
\begin{align}\label{eq:Kpurity}
\mathcal{P}\left(\rho_{\lvert\vec{\phi}\rvert \leq K}\right) \in \left[\frac{1}{K}, 1\right]
\end{align}
with minimum value $1/K$ occurring when all $K$ nonzero eigenvalues of $\rho_{\lvert\vec{\phi}\rvert \leq K}$ have equal value $1/K$. I.e. $\phi_{[k]} = 1/K\forall k$. Then, as maximally correlated states have $d$ or fewer nonzero eigenvalues,
\begin{align}\label{eq:rhoBounds}
\mathcal{P}(\rho_{MC}) \in \left[\frac{1}{d}, 1\right].
\end{align}

\section{Bounding entanglement}\label{sec:reeBound}
\subsection{Relative entropy of entanglement}\label{sec:ree}
The (quantum) relative entropy~\cite{nielsen2002quantum} of state $\rho$ to state $\sigma$ is 
\begin{align}
\begin{split}
	S(\rho || \sigma) = -S(\rho) - \tr(\rho \log_2 \sigma) \\
	\text{where } S(\rho) \equiv -\tr(\rho \log_2 \rho )
	\end{split}
\end{align}
and where $\log_2$ is the (base-2) matrix logarithm~\cite{Higham_logs}. Note that $S(\rho )$ is the von Neumann entropy of $\rho$~\cite{nielsen2002quantum}. Any quantum state $\sigma$ is Hermitian, positive semidefinite and thus diagonalizable with real, nonnegative eigenvalues. Let $U_\sigma$ be the matrix of eigenvectors for $\sigma$ (where each column is an eigenvector), then
\begin{align}
	\wedge_{\sigma} = U_\sigma^\dagger \sigma U_\sigma
\end{align}
where $\wedge_{\sigma}$ is the diagonal matrix of eigenvalues of $\sigma$ and $ U_\sigma^\dagger$ is the conjugate transpose of $U_\sigma$. For diagonalizable matrices, the matrix logarithm has the convenient property~\cite{Higham_logs}
\begin{align}
	\log_2 (\sigma) = U_\sigma \log_2 (\wedge_{\sigma}) U_\sigma^\dagger.
\end{align}
Note that the logarithm of a diagonal matrix is the diagonal matrix of the logarithm of each element.

Klein's inequality~\cite{nielsen2002quantum} shows $S(\rho || \sigma ) \geq 0$ for any quantum states $\rho$ and $\sigma$ with equality if and only if $\rho = \sigma$. Thus, the relative entropy roughly quantifies the distinguishability of two states. The relative entropy of entanglement is an entanglement measure defined
\begin{align}\label{eq:DreeDef}
\begin{split}
	D_{REE}(\rho ) &\equiv \min_\sigma S(\rho || \sigma) = S(\rho || \sigma^*) \\
	&= -S(\rho) - \tr(\rho \log_2 \sigma^*),
\end{split}
\end{align} 
where the minimum is taken over all separable states $\sigma$~\cite{PhysRevA.57.1619, Vedral_2002}. That is, $D_{REE}(\rho )$ is the relative entropy between $\rho$ and its most similar separable state $\sigma^*$.

\subsection{Bounds on relative entropy of entanglement for maximally correlated states}\label{sec:Dree_max}
We certify entanglement for maximally correlated systems $\rho_{MC}$ by lower-bounding the relative entropy of entanglement $D_{REE}(\rho_{MC})$ given knowledge of $\vec{\zeta}$ (Equation~\ref{eq:zetaDef}) and of the purity of $\rho_{MC}$:  $\mathcal{P}_{MC} =  \mathcal{P}(\rho_{MC})$ (Equation~\ref{eq:purity}). The experimental determination of these $d+1$ parameters is discussed in Section~\ref{sec:ExperimentalApplication}.

Maximally correlated states $\rho_{MC}$ of any dimension $d$ have the special property~\cite{ChenYang_2002} that the choice of separable $\sigma$ which minimizes $S(\rho_{MC} || \sigma)$ is always the diagonal system
\begin{align}\begin{split}
	\sigma^* 	&\equiv \sum_{k=1}^d \braket{k,k | \rho_{MC} | k,k} \ket{kk}\bra{kk} \\
				&= \sum_{k=1}^d \zeta_k \ket{kk}\bra{kk}.
\end{split}\end{align}

As maximally correlated $\rho_{MC}$ have $\braket{i,j | \rho_{MC} | k,l} = 0$ whenever $i\neq j$ or $k \neq l$,
\begin{align}\begin{split}
	\tr(\rho \log_2 \sigma^*) 	
								=	\sum_{k=1}^d \zeta_k \log_2(\zeta_k), \\ \text{ where } 0\cdot \log_2(0) \equiv 0.
\end{split}\end{align}

For any density matrix $\rho$ which is diagonalized by $U$ producing the diagonal matrix $\wedge$ of eigenvalues $ \vec{\lambda}$, i.e.
\begin{align*}
	\rho &= U \wedge U^\dagger ,
\end{align*}
it follows that
\begin{align}\begin{split}
-S(\rho) = \tr(\rho \log_2 \rho) 	&= 	\tr(\rho U \log_2 \wedge U^\dagger)\\
						&=	\tr(U^\dagger \rho U \log_2 \wedge ) \\
						&=	\tr(\wedge \log_2 \wedge ) \\
						&=	\sum_n \lambda_n \log_2 \lambda_n.
\end{split}\end{align}
Then if $\rho_{MC}$ has eigenvalues $\vec{\phi}$,
\begin{align}\label{eq:Dree}\begin{split}
	D_{REE} (\rho_{MC}) = \sum_k \phi_k \log_2 (\phi_k ) - \sum_k \zeta_k \log_2 (\zeta_k ), \\
	\text{ where } 0\cdot \log_2 (0) \equiv 0.
\end{split}\end{align}
Given $\vec{\zeta}$, the second sum on the RHS of Equation~\ref{eq:Dree} is computable. As we show in the following, the purity $\mathcal{P}_{MC}$ of $\rho_{MC}$ will serve to lower bound the first quantity (the negative von Neumann entropy $-S(\rho_{MC})$), i.e.
\begin{align}\label{eq:task}
	\sum_k \phi_k \log_2 (\phi_k ),
\end{align}
and thereby lower bound $D_{REE} (\rho_{MC})$. 

Both $\rho_{MC}$ and its eigenvalues $\vec{\phi}$ are unknown; however, the eigenvalues must obey Equation~\ref{eq:eigcons}, Equation~\ref{eq:puritycons}, and Equation~\ref{eq:majcons}. This information is sufficient to lower bound Equation~\ref{eq:task} leveraging the method of Lagrange multipliers. The method of Lagrange multipliers is an optimization strategy for finding local extrema of a function (such as Equation~\ref{eq:task}) subject to equality constraints (such as Equations~\ref{eq:eigcons} and \ref{eq:puritycons})~\cite{calcvar1966, de2000mathematical}.

\begin{figure}
	\centering
	\includegraphics[scale=1]{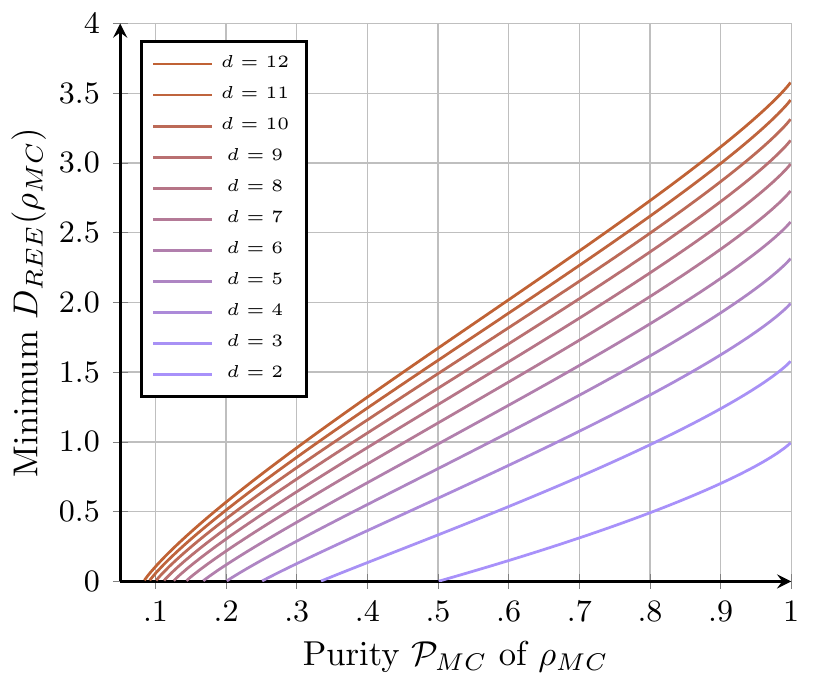}
	\caption{The lower bound for $D_{REE}(\rho_{MC})$ (Equation~\ref{eq:bound_mc}) as a function of measured purity $\mathcal{P}_{MC}$ for ideal $\rho_{MC}$ of dimensions $d=2$ through $d=12$. The $\rho_{MC}$ are ideal in that the are maximally correlated with $\braket{k,k | \rho | k,k} = 1/d \forall k$. Any other set of $\braket{k,k | \rho | k,k}$ will yield smaller lower bounds for a given $d$. Note that for $d=2$ and for $\mathcal{P}_{MC} = 1\forall d$ the equality holds.}
	\label{fig:Dree}
\end{figure}

The full derivation is lengthy and relegated to Appendix~\ref{appendix:lagrange}. We arrive at
\begin{widetext}
\begin{align}\label{eq:solution}
	-S(\rho_{MC}) = \sum_k \phi_k \log_2 (\phi_k )
		&\geq \phi_a \log_2(\phi_a) + (1 - \phi_a)\log_2\left(\frac{1 - \phi_a}{d - 1} \right),  \\
	 &\text{ where } \phi_a = \frac{1 + \sqrt{(d\mathcal{P}_{MC} - 1)(d - 1)}}{d}. \nonumber
\end{align}
Allowing the entropy of entanglement to be lower-bounded:
\begin{align} \label{eq:bound_mc}
	\begin{split}
	D_{REE}(\rho_{MC}) 	&= \sum_k \phi_k \log_2 (\phi_k ) - \sum_k \zeta_k \log_2 \zeta_k \\
						&\geq \phi_a \log_2 (\phi_a) + (1 - \phi_a) \log_2\left( \frac{1 - \phi_a}{d - 1} \right) - \sum_k \zeta_k \log_2 \zeta_k.
	\end{split}
\end{align}
\end{widetext}
for all $d \geq 2$, $\mathcal{P}_{MC} \in [1/d, 1]$. Figure~\ref{fig:Dree} plots the lower bound of Equation~\ref{eq:bound_mc} for equally maximally correlated systems: $\zeta_k = 1/d\forall k$ from $d=2$ to $d=12$-dimensional.

\subsection{Bounds for noisy maximally correlated states}\label{sec:D_noise}
Some level of noise is expected in any practical measurement. We assume the noise is white or incoherent and takes the form of a diagonal density operator,
\begin{align}
	\rho_{\text{noise}}  = \sum_{i} \sum_{j \neq i} \lambda_{i,j} \ket{i,j}\bra{i,j}
\end{align}
where entries $\braket{k,k |\rho_{\text{noise}} |k,k} = 0\forall k$. Then a maximally correlated system with noise takes the form
\begin{align}\label{eq:crho}
	\rho = \gamma \rho_{MC} + (1-\gamma ) \rho_{\text{noise}}. 
\end{align}
Noise present in terms $\braket{k,k | \rho | k,k}\forall k$ is not neglected; it is considered as a part of $\rho_{MC}$ rather than $\rho_{\text{noise}}$.

As $\rho_{MC}$ is maximally correlated, its $d$ eigenvectors $\ket{\phi_k}$ corresponding to nonzero eigenvalues $\phi_k$ must also be maximally correlated. Thus
\begin{align}
	\ket{\phi_k} = \sum_j \alpha_{j(k)}\ket{j,j}.
\end{align}
i.e.
\begin{align}
	\braket{i,j | \phi_k} = 0 \text{ } \forall i\neq j.
\end{align}
And so
\begin{align}
	\rho_{\text{noise}} \ket{\phi_k} = 0 \forall k.
\end{align}
The $d(d-1)$ eigenvectors of the diagonal $\rho_{\text{noise}}$ which correspond to the non-zero eigenvalues ($\{\lambda_{i,j}\}_{i\neq j}$) comprise the set $\{ \ket{i,j} \}_{i\neq j}$. We will henceforth denote this $d(d-1)$-length set $\{ \ket{m}\}$. We will write $\lambda_{i,j}$ as $\lambda_{m}$ and the entire $d(d-1)$-length array $\vec{\lambda}$. It follows,
\begin{align}
	\rho_{MC} \ket{m} = 0 \forall m.
\end{align}
Consequently
\begin{align}
	\begin{split}
	\rho \ket{m} 		&= \gamma \rho_{MC}\ket{m} + (1-\gamma ) \rho_{\text{noise}}\ket{m} \\
						&= 0 + (1-\gamma ) \lambda_{m}\ket{m}
	\end{split}\\
							\text{ and }& \nonumber \\
	\begin{split}
	\rho \ket{\phi_k} 	&= \gamma \rho_{MC}\ket{\phi_k} + (1-\gamma ) \rho_{\text{noise}}\ket{\phi_k} \\
						&= \gamma \phi_k \ket{\phi_k} + 0.
	\end{split}
\end{align}

The $d^2$ eigenvectors of $\rho$ are the $d$ eigenvectors $\{\ket{\phi_k}\}$ and the $d(d-1)$ eigenvectors $\{\ket{m}\}$ with eigenvalues $\gamma \phi_k$ and $(1-\gamma ) \lambda_{m}$. As a result, $\rho, \rho_{MC}$, and $\rho_{\text{noise}}$ are mutually diagonalizable, i.e.
\begin{subequations}
\begin{align}
	\wedge &= U^\dagger \rho U \\
	\wedge_{MC} &= U^\dagger \rho_{MC} U \\
	\wedge_\text{noise} &= U^\dagger \rho_{\text{noise}} U 
\end{align}
\end{subequations}
using the same $U$.

Donald's identity~\cite{donald_1987}: For density operator $\rho = \sum_k p_k \rho_k$ and density operator $\sigma$,
\begin{align}
\sum_k p_k S(\rho_k || \sigma) = S(\rho || \sigma) + \sum_k p_k S(\rho_k || \rho).
\end{align}
Let $\rho$ be that of Equation~\ref{eq:crho} and let $\sigma = \sigma^{**}$, an optimal separable state for which $D_{REE}(\rho ) = S(\rho || \sigma^{**})$. Then applying Donald's identity,
\begin{align}\label{eq:setup}
	\begin{split}
	S&(\rho || \sigma^{**}) = \\
		&\gamma S(\rho_{MC} || \sigma^{**}) + (1-\gamma ) S(\rho_{\text{noise}} || \sigma^{**}) \\
		&- \gamma S(\rho_{MC} || \rho) - (1 - \gamma )S(\rho_{\text{noise}} || \rho).
	\end{split}
\end{align}
As $\sigma^{**}$ is a separable state which may not be (in fact: is not) optimal for $\rho_{MC}$ or $\rho_{\text{noise}}$
\begin{align}
	S(\rho_{MC} || \sigma^{**}) 	&\geq	D_{REE}(\rho_{MC}) \\
	S(\rho_{\text{noise}} || \sigma^{**}) 	&\geq	D_{REE}(\rho_{\text{noise}}) = 0.
\end{align}

As for the third term in Equation~\ref{eq:setup},
\begin{align}
	S(&\rho_{MC} || \rho) = 
		\tr(\rho_{MC} \log_2 \rho_{MC} ) - \tr(\rho_{MC} \log_2 \rho) \nonumber \\
		&= \sum_k \phi_k \log_2(\phi_k) - \tr(\rho_{MC} U \log_2(\wedge)  U^\dagger) \nonumber \\
		&= \sum_k \phi_k \log_2(\phi_k) - \tr(U^\dagger \rho_{MC} U \log_2(\wedge))	\nonumber \\
		&= \sum_k \phi_k \log_2(\phi_k) - \tr(\wedge_{MC} \log_2(\wedge))	\\
		&= \sum_k \phi_k \log_2 (\phi_k) - \sum_k \phi_k \log_2(\gamma \phi_k) \nonumber \\
		&= \sum_k \phi_k \log_2 (1/\gamma) \nonumber \\
		&= -\log_2 (\gamma). \nonumber
\end{align}
And similarly for $\rho_{\text{noise}}$,
\begin{align}\begin{split}
	S(&\rho_{\text{noise}} || \rho) =\\
		&\tr(\rho_{\text{noise}} \log_2 \rho_{\text{noise}} ) - \tr(\rho_{\text{noise}} \log_2 \rho) \\
		&= -\log_2 (1-\gamma )
\end{split}\end{align}
Returning to Equation~\ref{eq:setup}
\begin{widetext}
\begin{align}\begin{split}\label{eq:noiseBound}
D_{REE}(\rho) &= S(\rho || \sigma^{**}) \\
		&= \gamma S(\rho_{MC} || \sigma^{**}) + (1-\gamma ) S(\rho_{\text{noise}} || \sigma^{**})
		- \gamma S(\rho_{MC} || \rho) - (1 - \gamma )S(\rho_{\text{noise}} || \rho) \\
		&= \gamma S(\rho_{MC} || \sigma^{**}) + (1-\gamma ) S(\rho_{\text{noise}} || \sigma^{**}) + \gamma\log_2(\gamma ) + (1 - \gamma )\log_2(1-\gamma) \\
		&\geq \gamma D_{REE}(\rho_{MC}) + \gamma \log_2(\gamma ) + (1 - \gamma ) \log_2(1-\gamma)
\end{split}\end{align}
\end{widetext}
The lower bound of $D_{REE}(\rho_{MC})$ from Section~\ref{sec:Dree_max} can be applied here; we need only determine the purity $\mathcal{P}_{MC}$ of $\rho_{MC}$ in order to apply Equation~\ref{eq:bound_mc}. $\mathcal{P}_{MC}$ can be determined from $\mathcal{P}$ of $\rho$:
\begin{align}\label{eq:purityconversion}
	\mathcal{P} &= \sum_k (\gamma \phi_k)^2 + \sum_m ((1-\gamma) \lambda_m)^2 \nonumber \\
				&= \gamma^2 \sum_k \phi_k^2 + \sum_m ((1-\gamma) \lambda_m)^2 \nonumber \\
				&= \gamma^2 \mathcal{P}_{MC} + (1-\gamma)^2 \sum_m (\lambda_m)^2 \nonumber \\
	\rightarrow \mathcal{P}_{MC} &= \frac{\mathcal{P} - (1-\gamma)^2 \sum_m (\lambda_m)^2}{\gamma^2}
\end{align}

In the incoherent noise case, we assume knowledge of each unique $\lambda_m$ in $\vec{\lambda}$ and apply Equation~\ref{eq:purityconversion}. In the white noise case, it is assumed $\lambda_m = \frac{1}{d(d-1)} \forall m$ and thus Equation~\ref{eq:purityconversion} becomes
\begin{align}\label{eq:purityconversion2}
	\mathcal{P}_{MC} &= \frac{\mathcal{P} - (1-\gamma)^2/(d(d-1))}{\gamma^2}.
\end{align}

\section{Experimental application} \label{sec:ExperimentalApplication}
For Equation~\ref{eq:bound_mc} and/or Equation~\ref{eq:noiseBound} to be applied, $\vec{\zeta}$, $\vec{\lambda}$, $\gamma$, and $\mathcal{P}$ must be experimentally measurable. We discuss measuring $\vec{\zeta}$ in the ideal noise-free case in Section~\ref{sec:ideal} and measuring $\vec{\zeta}$, $\vec{\lambda}$, and $\gamma$ for the noisy case in Section~\ref{sec:noise}. Measuring purity $\mathcal{P}(\rho)$ is discussed in Section~\ref{sec:Purity}.

\subsection{Maximally correlated states}\label{sec:ideal}
For biparite $d^2$-dimensional $\rho$ which is believed to be maximally correlated, a simultaneous measurement
\begin{align}
	\mathcal{q} = \sum_{i=1}^{d} \sum_{j\neq i} \braket{i,j | \rho | i,j}
\end{align}
of all non-correlated outcomes in the computational basis can be made. If $\mathcal{q} = 0$, $\rho$ is confirmed to be a maximally correlated state by definition (Equation~\ref{eq:MCDefinition}). Here, $\mathcal{q}$ is written as a single measurement, but practically may require $d$ distinct measurements which can then be summed. If, for instance, $\ket{i,j}_{AB}$ represents a two-photon state with photon $A$ in state $\ket{i}_A$ and photon $B$ in state $\ket{j}_B$, then we may first route photon $A$ state $\ket{1}_A$ to detector $a$ and photon $B$ states $\ket{2}_B$ through $\ket{d}_B$ to detector $b$ and measure simultaneous detection rate, determining $\sum_{j\neq 1} \braket{1,j | \rho | 1,j}$. The process may be iterated for $\ket{2}_A$ and $\ket{j\neq 2}_B$ and so on, until $\mathcal{q}$ can be computed from $d$ total measurements.

Once $\rho$ has been shown to be maximally correlated ($\rho = \rho_{MC}$) the $d$ distinct correlated measurements should be made,
\begin{align}
	\zeta_j = \braket{j,j | \rho_{MC} | j,j}
\end{align}
and purity measured (see Section~\ref{sec:Purity}). After these $2d + 1$ measurements,  Equation~\ref{eq:bound_mc} can be applied to lower bound $D_{REE}(\rho_{MC})$.

\subsection{Noisy maximally correlated states}\label{sec:noise}
If $\mathcal{q} \neq 0$ but is instead a small positive quantity -- and if the process which generates $\rho$ is a correlated one -- for example, a process which generates a bipartite system which must obey energy conservation and whose outputs therefore must be energy-correlated -- then we may treat $\rho$ as a maximally correlated state experiencing either white noise or incoherent noise, as in Equation~\ref{eq:crho}.

$\gamma$ can be determined, from the $d$ correlation measurements,
\begin{align}
	\gamma = \frac{\sum_{j=1}^{d} \braket{j,j | \rho | j,j}}{\mathcal{q} + \sum_{j=1}^{d} \braket{j,j | \rho | j,j}}.
\end{align}
Making
\begin{align}
	\zeta_j = \frac{\braket{j,j | \rho | j,j}}{\gamma}.
\end{align}
Depending on the system, we might either assume white noise,
\begin{align}\label{eq:lammequal}
	\lambda_{m} = \frac{\mathcal{q}}{d(d-1)},
\end{align}
or make the measurements of the uncorrelated states $\braket{i,j |\rho | i,j}$ more granular -- making up to $d(d-1)$ measurements in total -- to uniquely determine all $\lambda_{m}$. In either case, after these measurements are made, purity $\mathcal{P}$ should be measured (see Section~\ref{sec:Purity}) and then $\mathcal{P}_{MC}$ determined (using  Equation~\ref{eq:purityconversion} or Equation~\ref{eq:purityconversion2}). Then $D_{REE}(\rho_{MC})$ can be lower bounded using Equation~\ref{eq:bound_mc} and finally used in Equation~\ref{eq:noiseBound} to lower bound $D_{REE}(\rho)$.

In the white noise case, $1$ purity measurement is required, $d$ measurements are needed to determine all $d$ elements in $\braket{j,j |\rho |j,j}\forall j$, and $d$ measurements are used to determine $\mathcal{q}$. It is then assumed that all $d(d-1)$ elements in $\vec{\lambda}$ are equal (i.e. Equation~\ref{eq:lammequal}).  In the incoherent noise case, the same $1$ purity measurement and the same $d$ measurements are needed to determine $\braket{j,j |\rho |j,j}\forall j$. In this case we do not assume all elements in $\vec{\lambda}$ are equal and instead may choose to measure all $d(d-1)$ elements individually (i.e. $\braket{i,j |\rho |i,j}\forall i\neq j$). In summary in the white noise case, $2d + 1$ measurements are used. In the incoherent noise case, up to $d^2 + 1$ measurements are made.

\subsection{Measuring purity} \label{sec:Purity}
The purity of $\rho$ ($\mathcal{P}(\rho ) = \tr(\rho^2 )$) is a nonlinear quantity and cannot be directly measured from $\rho$. However, if two identically prepared $\rho$ can be simultaneously generated, then a direct linear measurement of the purity of $\rho$ is possible~\cite{Ekert2002_ShiftOpPurity}. For optical systems, the measurement can be made with a beamsplitter followed by a parity measurement~\cite{Ekert2002_ShiftOpPurity, islam2015measuring_GenHOM, PhysRevLett.109.020505, Moura2004_MultiPartiteEnt}. In this arrangement, the two-photon state $\rho$ is prepared twice, simultaneously (i.e. $\rho_1$ and $\rho_2$ are prepared where $\rho_1 = \rho_2 = \rho$). $\rho_1$ enters one input port of the beamsplitter and $\rho_2$ enters the other input port. The number of photons emerging from each port of the beamsplitter is recorded. A measurement is even-parity when the number of photons measured at each detector is even (i.e. 4-0, 0-4, or 2-2); a measurement is odd-parity when an odd number of photons are measured at each detector (i.e. 3-1, 1-3). A higher-order version of HOM interference~\cite{1987HOM} demands only even parity events occur when the states entering the beamsplitter are pure and identical, $\rho_1 = \rho_2 = \ket{\psi}\bra{\psi}$. When the incoming states are mixed or dissimilar, odd parity events will occur at an increasing rate. A parity measurement assigns a value of $x=+1$ to even-parity events and a value of $x=-1$ to odd-parity events. The expectation value $\braket{x} = \tr(\rho^2) = \mathcal{P}(\rho )$ returns the purity of $\rho$ (see Appendix~\ref{appendix:paritypurity} for a derivation of this result). While nontrivial, a direct measurement of purity is possible and may be preferable to the large number of interferometric measurements otherwise necessary for entanglement certification.

\section{Discussion}\label{sec:discussion}
\subsection{Entanglement dimension}
\begin{figure}
	\centering
	\includegraphics[scale=1]{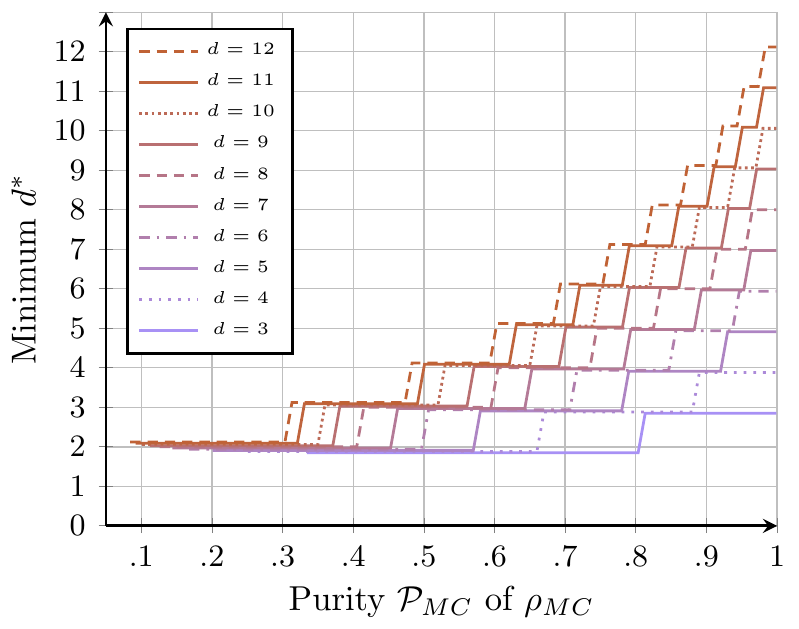}
	\caption{The lower bound for $d^*$ as a function of measured purity $\mathcal{P}_{MC}$ for noise-free $\rho_{MC}$ of dimensions $d=3$ through $d=12$ where $\braket{k,k | \rho_{MC} | k,k} = 1/d \forall k$. While the function is integer-valued, the plotted values have been offset somewhat for readability.}
	\label{fig:dstar}
\end{figure}

For a maximally-entangled bipartite state with subsystems of dimension $d$, the relative entropy of entanglement can be no greater than $\log_2 (d)$. Then for a given $D_{REE}^*$ the number of entangled dimensions in the system $d^*$ must be
\begin{align}
	d^* \geq \left\lceil 2^{D_{REE}^*}\right\rceil.
\end{align} 
This relationship allows us a rapid means of quantifying the dimensionality of entanglement in a high-dimensional system and providing an easily understood metric to users. In place of lower bounding $D_{REE}$ as in Figure~\ref{fig:Dree}, we can bound $d^*$. Figure~\ref{fig:dstar} bounds $d^*$ for ideal $\rho_{MC}$ of dimension $d=3$ to $d=12$ in the maximally correlated case where $\braket{k,k | \rho_{MC} | k,k} = 1/d \forall k$.

\subsection{Connection to mutual information} \label{sec:MI}
The majority of the work dedicated to bounding the relative entropy of entanglement in Section~\ref{sec:Dree_max} centered around lower bounding the negative von Neumann entropy $-S(\rho)$. Also appreciate that this quantity appears in the definition of the mutual information $I(\rho)$. Let $\rho$ be a two-party system comprised of party $A$ and party $B$.
\begin{align}\label{eq:MI}
	I(\rho) \equiv S(\rho^A) + S(\rho^B) - S(\rho).
\end{align}
Where $\rho^A = \tr_B(\rho)$ and $\rho^B = \tr_A(\rho)$. I.e. $\rho^A$ ($\rho^B$ ) is the two-party system $\rho$ with party $B$ ($A$) traced-out. $\tr_B$ ($\tr_A$) is a partial trace operation with respect to system $B$ ($A$). Assuming $\rho$ is a maximally correlated system, $\rho_{MC}$:
\begin{align}
	\rho_{MC}^A 
		&= \tr_B (\rho_{MC}) \nonumber \\
		&= \sum_k \bra{k}_B \sum_{u,v} \alpha_{u,v} \ket{u,u}_{AB}\bra{v,v}_{AB} \ket{k}_B \nonumber \\
		&= \sum_k \alpha_{k,k} \ket{k}_A\bra{k}_A.
\end{align}
And similarly, $\rho_{MC}^B = \sum_k \alpha_{k,k} \ket{k}_B\bra{k}_B$. That is, both $\rho_{MC}^A$ and $\rho_{MC}^B$ are identical, diagonal operators. By Equation~\ref{eq:zetaDef}, $\alpha_{k,k} = \zeta_k$. Therefore
\begin{align}
	S(\rho_{MC}^A) = S(\rho_{MC}^B) = -\sum_k \zeta_k \log_2 \zeta_k.
\end{align}
Inserting into Equation~\ref{eq:MI},
\begin{align}\label{eq:MI2}
	I(\rho_{MC})
		&= S(\rho_{MC}^A) + S(\rho_{MC}^B) - S(\rho_{MC}) \nonumber \\
		&= -2\sum_k \zeta_k \log_2 \zeta_k + -S(\rho_{MC}).
\end{align}
Therefore, for maximally correlated systems, the lower bound of $-S(\rho_{MC})$ found in this work (Equation~\ref{eq:solution}) also enables the lower bound of quantities such as mutual information (Equation~\ref{eq:MI}). With some modifications, mutual information can also be lower-bounded for the noisy maximally correlated states like those covered in Section~\ref{sec:D_noise}.

\subsection{Von Neumann entropy in other measures}
Taking a broader perspective: while this work is motivated by the certification of entanglement in high-dimensional, maximally-correlated states, the lower bound derived in Appendix~\ref{appendix:lagrange} (Equation~\ref{eq:solution}/Equation~\ref{eq:apFinal}) has greater applicability. Equation~\ref{eq:apFinal} is the lower bound on the negative von Neumann entropy of a state, $-S(\rho )$, found using Equation~\ref{eq:eigcons} and Equation~\ref{eq:puritycons}. This state $\rho$ need not be maximally correlated. If $K$ is the number of nonzero eigenvalues in state $\rho$ and $\mathcal{P}$ is the purity of $\rho$ then Equation~\ref{eq:apFinal} still holds. Letting $K$ replace $d$ and $\mathcal{P}$ replace $\mathcal{P}_{MC}$:
\begin{align}\label{eq:general}
	\begin{split}
	-S(&\rho) \geq \\
	&\phi_a \log_2(\phi_a) + (1 - \phi_a)\log_2\left(\frac{1 - \phi_a}{K - 1} \right),
	\end{split} \\
	&\text{where } \phi_a = \frac{1 + \sqrt{(K\mathcal{P} - 1)(K - 1)}}{K}. \nonumber
\end{align}
With knowledge of purity $\mathcal{P}$ and $K$ for a state $\rho$, this lower bound on the negative von Neumann entropy may prove useful for other quantum information applications. Negative von Neumann entropy, in addition to appearing in the expression for mutual information (see Section~\ref{sec:MI}), also appears in the super dense coding capacity of a channel~\cite{Sazim2013}. (Positive) von Neumann entropy appears in other information applications~\cite{PhysRevLett.122.210402}, which may find the upper bound of this quantity useful.

As for the two required parameters for an arbitrary state $\rho$: Purity $\mathcal{P}$ can be measured as described in Section~\ref{sec:Purity}. As shown in Section~\ref{sec:Schur}, the maximum number of nonzero eigenvalues $K$ is less or equal to the number of nonzero measurements of $\braket{i,j | \rho | i,j}\forall i \forall j$. See Section~\ref{sec:D_noise} for a discussion of this measurement. Appreciate, the bounds on Equation~\ref{eq:general} become more favorable for smaller values of $K$. These two parameters, $\mathcal{P}$ and $K$, are experimentally accessible for arbitrary $\rho$ with $d^2 + 1$ measurements.

\section{Conclusion}
There exists a fundamental trade-off for separable and dimension-limited states between their strength of correlations in any basis and their purity. Here, we have applied this trade-off and derived useful lower bounds for system entanglement using the relative entropy of entanglement for maximally-correlated quantum states. These bounds allow experimental entanglement certification for high-dimensional systems while demanding neither a large number of measurements in multiple (unbiased) bases of a quantum state nor a large number of interferometric measurements on a quantum state. The challenges of both of these approaches have been replaced by a single difficult measurement: the quantum system's purity. Purity itself is not even required: only a valid lower bound. While such a measurement is challenging in its own right, it is possible to make the measurement directly and significantly reduces the number of difficult measurements needed to certify a high-dimensional system's entanglement.

Avenues for improvement on this work include fewer assumptions regarding the form of the noise on the system and potentially including bounds for non-maximally correlated states.

\section{Acknowledgments}
We thank Ruichao Ma for discussion around mutual information. We would like to acknowledge the financial support by the National Science Foundation (NSF) under grant 1839191-ECCS.

\printbibliography

\onecolumn
\appendix
\section{Definition of maximally correlated states}\label{appendix:mc}
\beginappendix{A}
We show that the definition of maximally correlated states in Ref~\cite{Hiroshima2004_SchmidtCorrelatedDefinition} -- that any maximally correlated state $\rho_{MC}$ can be written
\begin{align}
	\rho_{MC}= \sum_{u,v=1}^{\min \{ d_A, d_B\}} \alpha_{uv} \ket{uu}_{AB} \bra{vv}_{AB}
\end{align}
where $d_A$ and $d_B$ are the dimensions of subsystems $A$ and $B$ respectively -- is equivalent to the one used in this work --
\begin{align}
 \braket{i,j|\rho_{MC} | i,j}=0 \forall i\neq j.
\end{align}
-- when $\rho_{MC}$ is a density operator.

In the forward direction
\begin{align}\begin{split}
	\braket{i,j | \rho_{MC} | i,j}
		&= \braket{i,j |  \sum_{u,v=1}^{\min \{ d_A, d_B\}} \alpha_{uv} \ket{u,u} \bra{v,v} |i,j}\\
		&= \sum_{u,v=1}^{\min \{ d_A, d_B\}} \alpha_{uv} \braket{i,j  | u,u} \braket{v,v |i,j}\\
		&= 
		\begin{cases} 
      \alpha_{j,j}, & i=j \\
      0, & i\neq j
   \end{cases}\\
   \rightarrow \braket{i,j | \rho_{MC} | i,j} &= 0 \forall i\neq j.
\end{split}\end{align}

As for the reverse direction: Let $\rho_{MC}$ be a density operator, meaning it can be written
\begin{align}
	\rho_{MC} =\sum_u w_u \ket{\psi_u}\bra{\psi_u}
\end{align}
where $w_u$ are all positive real. Then
\begin{align}\begin{split}
	\braket{i,j|\rho_{MC} | i,j}&=0 \forall i\neq j\\
	\sum_u w_u \lvert\braket{i,j |\psi_u}\rvert^2 &= 0 \forall i\neq j\\
	\text{As $w_u$ are positive}, \rightarrow  \braket{i,j |\psi_u}&=0 \forall u \forall i\neq j.\\
	\text{therefore, } \braket{i,j|\rho | k,l} = \sum_u w_u \braket{i,j |\psi_u}\braket{\psi_u | k,l} &= 0\text{ when } i\neq j \text{ or } k\neq l.\\
\end{split}\end{align}
Thus, $\braket{i,j|\rho | k,l}$ can take a potentially non-zero value $\alpha_{jk}$ only when $i = j$ and $k = l$,
\begin{align}\begin{split}
	\braket{j,j|\rho_{MC} | k,k} = \alpha_{jk}\forall{j,k}
	\rightarrow \rho_{MC}	=	\sum_{u,v=1}^{\min \{d_A, d_B \}} \alpha_{uv} \ket{u,u}\bra{v,v}
\end{split}\end{align} 

As both definitions imply one another for any density operator $\rho_{MC}$, they are equivalent.

\section{Lower bound via method of Lagrange multipliers}\label{appendix:lagrange}
\beginappendix{B}
Section~\ref{sec:Dree_max} observes that -- for maximally correlated systems -- a lower bound on the relative entropy of entanglement can be determined by finding the array of nonzero eigenvalues $\vec{\phi}$ which minimize Equation~\ref{eq:task} (the negative von Neumann entropy) while obeying Equation~\ref{eq:eigcons}, Equation~\ref{eq:puritycons}, and Equation~\ref{eq:majcons}. Repeated here, with the bounds made explicit:
\begin{align}
	&\sum_{k=1}^d \phi_k \log_2 (\phi_k )			\tag{\ref{eq:task}}			\\
	&\sum_{k=1}^d \phi_k = 1						\tag{\ref{eq:eigcons}}		\\
	&\mathcal{P}_{MC} = \sum_{k=1}^d \phi_k^2		\tag{\ref{eq:puritycons}}	\\
	&\sum_{k=1}^m \phi_{[k]} \geq \sum_{k=1}^m \zeta_{[k]}, \text{ for } m=1, 2, 3, ... d \text{ with equality for } m=d \tag{\ref{eq:majcons}}
\end{align}
where $\mathcal{P}_{MC} \in [1/d, 1]$ (Equation~\ref{eq:rhoBounds}) and $\vec{\zeta}$ are known quantities.

Before preceding to the method of Lagrange multipliers, which can be used to lower bound Equation~\ref{eq:task} given the constraints, we note the cases where the constrained solution is unique and thus no lower bound need be found. For $\mathcal{P}_{MC} = 1$, there is only one possible set of eigenvalues which satisfy both Equation~\ref{eq:eigcons} and Equation~\ref{eq:puritycons}: $\phi_k = \delta_{1,k}\forall k$. In this case, Equation~\ref{eq:task} $= 0$. For $\mathcal{P}_{MC} = 1/d$, the only possible set of eigenvalues is $\phi_k = 1/d \forall k$. In this case, Equation~\ref{eq:task} $= -\log_2(d)$. Finally, the constraint equations also yield a unique solution for all $\mathcal{P}_{MC}$ when $d=2$. In this case the two eigenvalues, $\phi_{+}$ and $\phi_{-}$, are $\phi_{\pm} = .5 \pm .5\sqrt{2\mathcal{P}_{MC} - 1}$. In conclusion, determining the minimum constrained value of Equation~\ref{eq:task} is nontrivial only when $d\geq 3$ and $\mathcal{P}_{MC} \in (1/d, 1)$. The nontrivial cases will necessitate using the method of Lagrange multipliers.

Again, the method of Lagrange multipliers is an optimization strategy for finding local extrema of a function (such as Equation~\ref{eq:task}) subject to equality constraints (such as Equations~\ref{eq:eigcons} and \ref{eq:puritycons}). The method finds the array of parameters $\vec{\phi^*}$ (subject to the equality constraints) which either maximizes or minimizes the function of interest, Equation~\ref{eq:task}.  That array can then be substituted into the function of interest (the objective function) to return its constrained extremum. When this extremum is a minimum, it will serve as a lower bound for Equation~\ref{eq:task}.

Derivation of the constrained minimum will proceed with four major steps. First, the function given by Equation~\ref{eq:task} (a function of the $d$-length array of eigenvalues, $\vec{\phi} = \braket{\phi_1, \phi_2, ..., \phi_d}$) is not fully compatible with the method of Lagrange multipliers. We will replace it with a function $f(\vec{\phi})$, which approximates Equation~\ref{eq:task} with precision determined by $\epsilon$. As $\epsilon\rightarrow 0$, $f(\vec{\phi})$ goes to Equation~\ref{eq:task}. (At the conclusion of the derivation, we will find the constrained minimum of $f(\vec{\phi})$ is independent of $\epsilon$ once $\epsilon$ is sufficiently close to zero. Therefore the minimum also applies to Equation~\ref{eq:task}). Second: the method of Lagrange multipliers will be used on the approximate function $f(\vec{\phi})$. It will not definitively return the optimal $\vec{\phi^*}$, but rather will show us the form the optimal solution must take. Third: it will follow from step two that for any given $d\geq 3$ and any given $\mathcal{P}_{MC} \in (1/d, 1)$, we will find $d-1$ candidate solutions, one of which will be the optimal $\vec{\phi^*}$. Function $f(\vec{\phi})$ will be written in terms of these candidate solutions, simplifying the expression and allowing it to be written as a function of a single discrete variable $s_a$: $f(s_a)$. This will prepare us for the fourth and final step: determining which of these candidates minimizes $f(s_a)$; this delivers us $\vec{\phi^*}$ and the constrained minimum $f(\vec{\phi} = \vec{\phi^*})$ and thus the lower bound of Equation~\ref{eq:task}.

\subsection{Approximate function}
Applying the method of Lagrange multipliers generally requires that the objective function and the constraint function(s) have continuous first derivatives (and continuous second derivatives are also frequently assumed). In addition, the method requires that the entire domain $\mathbb{R}^d$ is considered (where $d$ is the size of $\vec{\phi }$). The (negative) entropy function $\phi\log(\phi)$ suffers from a lack of continuity at $\phi =0$ and is not defined for $\phi <0$. To sidestep these issues, we define
\begin{align}
	f_1(\phi) = 
	\begin{cases} 
      \phi\log_2(\phi) 										&	\phi > \epsilon \\
       \frac{\phi^2}{\epsilon 2\ln(2)} + \phi\log_2(\epsilon) - \frac{\epsilon}{2\ln(2)} 		&	\phi \leq \epsilon
   \end{cases}
\end{align}
for some small $\epsilon$ where $0 < \epsilon \ll 1/d$.

The constructed function $f_1(\phi )$ is continuous with continuous first and second derivatives:
\begin{align}
	\frac{df_1(\phi )}{d \phi} &= 
	\begin{cases} 
    	\frac{1}{\ln(2)}  + \log_2(\phi )						& \phi > \epsilon \\
      	\frac{\phi }{\epsilon \ln(2)} + \log_2(\epsilon)		& \phi \leq \epsilon 
   	\end{cases}; \\
	\frac{d^2 f_1(\phi )}{d\phi^2} &= 
	\begin{cases} 
    	\frac{1}{\phi\ln(2)}			 						& \phi > \epsilon \\
      	\frac{1}{\epsilon \ln(2)}	 							& \phi \leq \epsilon
   	\end{cases}.
\end{align}
$f_1(\phi)$ is equal to the (negative) entropy function $\phi\log_2(\phi )$ for $\phi > \epsilon$ and approximates it for $0 \leq \phi \leq \epsilon$, with the greatest deviation occurring at $\phi=0$ where $0\cdot\log_2(0) \equiv 0$ and $f_1(0) = \frac{-\epsilon}{2\ln(2)}$. The (negative) entropy function is undefined for $\phi <0$. Function $f_1$, however, is defined everywhere in $\mathbb{R}$. Instead of working with Equation~\ref{eq:task}, we leverage $f_1(\phi )$ and define the replacement objective function
\begin{align}\label{eq:newobj}
	f(\vec{\phi}) = \sum_{k=1}^d f_1(\phi_k)
\end{align}
where $\vec{\phi} = \braket{\phi_1, \phi_2, ... \phi_d} \in \mathbb{R}^d$. Function $f(\vec{\phi})$ maps an array of $d$ real numbers to a single real value ($f: \mathbb{R}^d \rightarrow \mathbb{R}$). This concludes the first step in our derivation, finding a function which can approximate Equation~\ref{eq:task} and is compatible with the method of Lagrange multipliers. We now proceed to the second step: the actual application of the method of Lagrange multipliers.

\subsection{Method of Lagrange multipliers}
We wish to minimize $f(\vec{\phi})$ (Equation~\ref{eq:newobj}), given the constraints
\begin{align}\label{eq:g1}
	g_1(\vec{\phi}) &= \left( \sum_{k=1}^d \phi_k\right) - 1 = 0\\
	\label{eq:g2}
	g_2(\vec{\phi}) &= \left( \sum_{k=1}^d \phi_k^2\right) - \mathcal{P}_{MC} = 0,
\end{align}
which are a reformat of Equations~\ref{eq:eigcons} and \ref{eq:puritycons} where $\mathcal{P}_{MC}$ is in $(1/d, 1)$.

Let $\vec{\phi^*}$ be an input which minimizes $f(\vec{\phi})$ while satisfying the constraints, i.e.
\begin{align}
	f(\vec{\phi^*}) = \min_{\vec{\phi}\in\mathbb{R}^d \text{ s.t. } g_1(\vec{\phi}) = g_2(\vec{\phi}) = 0}\left( f(\vec{\phi}) \right)
\end{align}
(where s.t. stands for ``subject to''). As $f(\vec{\phi})$, $g_1(\vec{\phi})$, and $g_2(\vec{\phi})$ all have continuous first (and second) derivatives everywhere: by the Lagrange multiplier theorem~\cite[Chap~1, Sect~9]{calcvar1966}~\cite[Chap~7, Sect~1]{de2000mathematical}, there will exist unique and real constants $\lambda_1$, $\lambda_2$ such that 
\begin{align}\label{eq:LM}
	\nabla f(\vec{\phi^*}) = \lambda_1 \nabla g_1(\vec{\phi^*}) + \lambda_2 \nabla g_2(\vec{\phi^*}).
\end{align}
where $\nabla f(\vec{\phi^*})$ is the length-$d$ vector of partial derivatives of $f(\vec{\phi})$ evaluated at $\vec{\phi^*}$, i.e. $\braket{\frac{\partial f(\vec{\phi})}{\partial \phi_k}|_{\phi^*_k}}$. $\nabla g_n(\vec{\phi^*})$ for $n=1,2$ are similarly defined. The Lagrange multiplier theorem will hold so long as the vectors $\nabla g_1(\vec{\phi^*})$ and $\nabla g_2(\vec{\phi^*})$ are linearly independent at $\vec{\phi^*}$ and so long as we are considering the case where $\vec{\phi}$ is at least length-3, i.e. $d\geq 3$, (as the number of variables must exceed the number (two) of constraints). 

Equation~\ref{eq:LM} can be written as $d$ equations (for $k=1,2,..., d$). For each $k$:
\begin{align}\label{eq:allk}
	\begin{cases}
		\frac{1}{\ln(2)} + \log_2(\phi^*_k) = \lambda_1 + 2 \phi^*_k\lambda_2, 						& \phi^*_k > \epsilon \\
		\frac{\phi^*_k}{\epsilon \ln(2)} + \log_2(\epsilon)  = \lambda_1 + 2 \phi^*_k\lambda_2,		& \phi^*_k \leq \epsilon		
	\end{cases} 
\end{align}

We use Equation(s)~\ref{eq:allk} to examine the form the optimal solution $\vec{\phi^*}$ will take.  If $\vec{\phi^*}$ contains multiple elements less than $\epsilon$ ($\phi^*_n, \phi^*_m \leq \epsilon$) then
\begin{align}
	\lambda_1 	= \frac{\phi^*_n}{\epsilon \ln(2)} + \log_2(\epsilon) - 2 \phi^*_n\lambda_2 \nonumber
				&= \frac{\phi^*_m}{\epsilon \ln(2)} + \log_2(\epsilon) - 2 \phi^*_m\lambda_2 \nonumber \\
	\rightarrow	\phi^*_n\left(\frac{1}{\epsilon \ln(2)} - 2\lambda_2  \right)
				&= \phi^*_m\left(\frac{1}{\epsilon \ln(2)} - 2\lambda_2  \right)\nonumber \\
	\rightarrow \phi^*_n &= \phi^*_m \text{ for all } \phi^*_n, \phi^*_m \leq \epsilon
\end{align}
Thus if the optimal solution $\vec{\phi^*}$ contains multiple elements $\leq\epsilon$ all such elements are equal. We call this value $\phi_s \leq \epsilon$. That is: if $\phi^*_k \leq \epsilon$ then $\phi^*_k = \phi_s$.

If $\vec{\phi^*}$ contains multiple elements greater than $\epsilon$  ($\phi^*_n, \phi^*_m > \epsilon$) then
\begin{align}\label{eq:testlam2}
	\lambda_1	&=	\frac{1}{\ln(2)} + \log_2(\phi^*_n)  -  2 \phi^*_n\lambda_2
				=	\frac{1}{\ln(2)} + \log_2(\phi^*_m)  -  2 \phi^*_m\lambda_2                                                                                                                                                                                                                                                                                                                                                                                                                                                                                                                                                                                                                                                                                                                                                                                                                                                                                                                                                                                                                                                                        \nonumber \\
	\rightarrow \lambda_2	&=	\frac{\log_2 (\phi^*_m / \phi^*_n)}{2(\phi^*_m - \phi^*_n)} \text{ when } \phi^*_m \neq \phi^*_n \text{ and } \phi^*_n, \phi^*_m > \epsilon .
\end{align}
Note the function on the R.H.S. of Equation~\ref{eq:testlam2} is, for a given $\phi^*_n$, a monotonically decreasing function of $\phi^*_m$ (ignoring the point where $\phi^*_m = \phi^*_n$). In order for $\lambda_2$ to be a constant, only two (or fewer) distinct values are permitted for all $\phi^*_k \in \vec{\phi^*}$ where $\phi^*_k \geq \epsilon$. We call these values $\phi_a$ and $\phi_b$ (where $\phi_a > \phi_b$). That is, for all $k$, if $\phi^*_k \geq \epsilon$ then either $\phi^*_k = \phi_a $ or $\phi^*_k = \phi_b$. When $\vec{\phi^*}$ contains only one unique value $> \epsilon$, we will label it $\phi_a$ and leave quantity $\phi_b$ undefined.

Thus far, we see the optimal solution $\vec{\phi^*}$ contains three or fewer unique elements: $\phi_a > \phi_b > \epsilon \geq \phi_s$ where again $\epsilon$ is a positive quantity taken arbitrarily close to zero. In fact, in the following we will show $\vec{\phi^*}$ contains only two or fewer unique elements.

To meet constraint $g_1$ (Equation~\ref{eq:g1}), $\vec{\phi^*}$ must have nonnegative elements of appreciable magnitude ($> \epsilon$). Consequently, $\vec{\phi^*}$ must contain some elements equal to $\phi_a$ and may or may not contain other elements equal to $\phi_b$ and/or $\phi_s$. If $\vec{\phi^*}$ contains both element $\phi_\mathcal{A} > \epsilon$ (where $\mathcal{A}=a$ or $\mathcal{A}=b$) and element $\phi_s$ then
\begin{align}\label{eq:testlam2s}
	\lambda_1 	&=	\frac{1}{\ln(2)} + \log_2(\phi_\mathcal{A})  -  2 \phi_\mathcal{A}\lambda_2 
				=	\frac{\phi_s}{\epsilon \ln(2)} + \log_2(\epsilon) - 2 \phi_s\lambda_2	\nonumber \\
	\rightarrow \lambda_2	&=	\frac{\log_2(\phi_\mathcal{A} / \epsilon) + (1 - \frac{\phi_s}{\epsilon})\log_2(e)}{2(\phi_\mathcal{A} - \phi_s )}.
\end{align}
where $e$ is Euler's number. We will show presently that, for a given $\phi_s$, the function on the R.H.S of Equation~\ref{eq:testlam2s} is a monotonically decreasing function of $\phi_\mathcal{A}$ (whenever $\phi_\mathcal{A} > \epsilon \geq \phi_s$). As $\lambda_2$ is a constant, only one distinct value of $\phi_\mathcal{A}$ is permitted. Therefore, if the optimal solution  $\vec{\phi^*}$ contains element(s) equal to $\phi_s$ then all other elements are equal to $\phi_a$ and no third value (represented by $\phi_b$) can appear in $\vec{\phi^*}$. The R.H.S of Equation~\ref{eq:testlam2s} is shown to be monotonic as follows: See that the derivative of the R.H.S. is
\begin{align}\label{eq:dRHS}
	\frac{\partial (\text{R.H.S. of Equation~\ref{eq:testlam2s}} )}{\partial \phi_\mathcal{A}}
	&= \frac{\log_2(e)}{2(\phi_\mathcal{A} - \phi_s)^2} \cdot \left(
		\ln(\epsilon / \phi_\mathcal{A})  + \phi_s \left(\frac{1}{\epsilon} - \frac{1}{\phi_\mathcal{A}}\right)
		\right).
\end{align}
Of course, the multiplier term $\frac{\log_2(e)}{2(\phi_\mathcal{A} - \phi_s)^2}$ is positive and the first term $\ln(\epsilon / \phi_\mathcal{A})$ is negative. As $\epsilon^{-1} - \phi_\mathcal{A}^{-1} > 0$, the second term is nonpositive when $\phi_s \leq 0$. Therefore Equation~\ref{eq:dRHS} is negative when $\phi_s \leq 0$.

Now we consider the case where $\phi_s$ is positive, i.e. $\phi_s \in (0, \epsilon]$. It follows
\begin{align}\label{eq:logderiv}
	\phi_s \left(\frac{1}{\epsilon} - \frac{1}{\phi_\mathcal{A}}\right) < 1 - \frac{\epsilon}{\phi_\mathcal{A}}.
\end{align}
Note that
\begin{align}\label{eq:logx}
	1 - x < \ln(1/x) \text{  } \forall x\in (0,1).
\end{align}
As $\phi_\mathcal{A} > \epsilon$, substitute $x= \epsilon / \phi_\mathcal{A}$ and apply Equation~\ref{eq:logx} to the R.H.S. of Equation~\ref{eq:logderiv}:
\begin{align}
	1 - \frac{\epsilon}{\phi_\mathcal{A}} < \ln\left(\frac{\phi_\mathcal{A}}{\epsilon}\right) = \left\lvert \ln\left(\frac{\epsilon}{\phi_\mathcal{A}}\right)  \right\rvert
\end{align}
then
\begin{align}
	\phi_s \left(\frac{1}{\epsilon} - \frac{1}{\phi_\mathcal{A}}\right) < \left\lvert \ln\left(\frac{\epsilon}{\phi_\mathcal{A}}\right)  \right\rvert
\end{align}
when $\phi_s$ is positive.  As the magnitude of the negative term ($\ln(\epsilon / \phi_\mathcal{A})$) exceeds that of the positive term ($\phi_s(\epsilon^{-1} - \phi_\mathcal{A}^{-1})$) in Equation~\ref{eq:dRHS}, the quantity is negative (for $\phi_s \in (0, \epsilon ]$).

Having shown Equation~\ref{eq:dRHS} is negative for both positive and negative $\phi_s$, we can state generally that
\begin{align}
	\frac{\partial (\text{R.H.S. of Equation~\ref{eq:testlam2s}} )}{\partial \phi_\mathcal{A}} < 0 \text{ for } \phi_\mathcal{A} > \epsilon \geq \phi_s,
\end{align}
verifying the initial claim that R.H.S. of Equation~\ref{eq:testlam2s} is a monotonically decreasing function of $\phi_\mathcal{A}$ (whenever $\phi_\mathcal{A} > \epsilon$). The significance of this result: if the optimal solution contains elements $\phi_a$ and $\phi_s$ it cannot also contain value $\phi_b$ because -- as $\lambda_2$ (Equation~\ref{eq:testlam2s}) is a constant -- two distinct values of $\phi_\mathcal{A}$ (e.g. both $\phi_a$ and $\phi_b$) cannot satisfy it.

In summary, as the optimal solution $\vec{\phi^*}$  must contain element(s) equal to $\phi_a$ and cannot simultaneously contain both $\phi_s$ and $\phi_b$: when the method of Lagrange multipliers holds, the optimal solution $\vec{\phi^*}$ contains two or fewer unique elements. We call these elements $\phi_a$ and $\phi_x$. As previously defined, $\phi_a$ must exist and be strictly greater than $\epsilon$. If it exists, $\phi_x$ is less than $\phi_a$ and can be either greater or less than $\epsilon$. However, as a consequence of constraint $g_2$ (Equation~\ref{eq:g2})
\begin{align}\label{eq:pm1}
	\left\lvert \phi^*_k\right\rvert \leq \sqrt{\mathcal{P}_{MC}} < 1 \forall k \nonumber \\
	\rightarrow \phi_a, \left\lvert \phi_x \right\rvert < 1.
\end{align}
and therefore $\phi_a \in (\epsilon, 1)$ and $\phi_x \in (-1, \phi_a)$.

The method of Lagrange multipliers can only fail to find the optimal constrained solution when $\nabla g_1(\vec{\phi^*}) \propto \nabla g_2(\vec{\phi^*})$. This occurs only when all elements in $\vec{\phi^*}$ are equal. I.e. this is the case where $\phi_x$ does not exist and $\phi_k^* = \phi_a = 1/d\forall k$. This case is trivial and (by the constraints) only occurs when $\mathcal{P}_{MC} = 1/d$. As we are interested in the region $\mathcal{P}_{MC} \in (1/d, 1)$, this trivial case can never occur and therefore there are no points where the method of Lagrange multipliers can fail. If an optimal solution exists (when purity $1/d < \mathcal{P}_{MC} < 1$ and dimension $d \geq 3$), the method of Lagrange multiplier cannot fail to find it.

An optimal solution does exist. The function $f_1(\phi)$ has a minimum value ($f_1(\phi = e^{-1}) = -\log_2(e)/e$) and consequently $f(\vec{\phi})$ cannot decrease indefinitely. As $f(\vec{\phi})$ is clearly not a constant, it has a constrained minimum. As this minimum does not occur in a region where the method of Lagrange multipliers can fail (where the elements of $\vec{\phi}$ are all equal), it will be identified by the method of Lagrange multipliers.

This completes the second step. We have determined that the constrained minimum of $f(\vec{\phi})$ occurs at a $\vec{\phi}$ of the form
\begin{align}\label{eq:phistar}
	\vec{\phi^*} = \braket{\phi_1 = \phi_a, \phi_2 = \phi_a, ..., \phi_{s_a} = \phi_a, \phi_{s_a+1} = \phi_x, ..., \phi_d = \phi_x }
\end{align}
(and all permutations thereof; the function is symmetric) where $s_a$ is some integer index less or equal to $d-1$. In the third step the function $f(\vec{\phi})$ will be written in terms of $\phi_a$ and $\phi_x$ and ultimately in terms of $s_a$. 

\subsection{Candidate solutions}
As the optimal solution $\vec{\phi^*}$ contains two unique values $\phi_a > \phi_x$, the constraints (Equation~\ref{eq:g1} and Equation~\ref{eq:g2}) can be written
	\begin{align}
	s_a \phi_a + (d - s_a) \phi_x  &= 1 \label{eq:sa1}\\
	s_a \phi_a^2 + (d - s_a) \phi_x^2 &= \mathcal{P}_{MC} \label{eq:sa2}.
\end{align}
where $s_a$ is an integer between $1$ and $d-1$ and where $1 > \phi_a \gg \epsilon$ and $\phi_x \in (-1 , \phi_a )$. (The $\pm 1$ limits are a consequence of Equation~\ref{eq:pm1}.) Each value of $s_a$ corresponds to a different candidate solution, one of which (by the method of Lagrange multipliers) will be optimal.

Appreciate that the average value of the $d$ terms in Equation~\ref{eq:sa1} is $1/d$. As $\phi_a$ and $\phi_x$ are both present and unique and as $\phi_a$ is the larger of the two: $\phi_a > 1/d$ and $\phi_x < 1/d$. This allows us to select the proper root when solving for Equation~\ref{eq:sa1} and Equation~\ref{eq:sa2} to arrive at
\begin{align}
	\phi_a	&= \frac{1 + \sqrt{\frac{d - s_a}{s_a}\cdot \left(d\mathcal{P}_{MC} - 1\right) }   }{d}
		=\frac{1 - (d - s_a)\phi_x}{s_a},\label{eq:ab_a}\\
	\phi_x	&= \frac{1 - \sqrt{\frac{s_a}{d - s_a}\cdot \left(d\mathcal{P}_{MC} - 1\right) }   }{d}
		=\frac{1 - s_a\phi_a}{d - s_a}. \label{eq:ab_b}
\end{align}

As an aside, it follows
\begin{align}
	\phi_x - \phi_a = \frac{-\sqrt{d\mathcal{P}_{MC} - 1}}{\sqrt{s_a(d-s_a)}}.
\end{align}
And as $s_a \in[1, d-1]$ and $\mathcal{P}_{MC} \in (1/d, 1)$, the numerator $-\sqrt{d\mathcal{P}_{MC} - 1}\in (0, -\sqrt{d-1})$ and the denominator \\
$\sqrt{s_a(d-s_a)}\in [\sqrt{d-1}, d/2]$. Consequently:
\begin{align}\label{eq:axdifference}
	-1 = \frac{-\sqrt{d-1}}{\sqrt{d-1}} < \phi_x - \phi_a < \frac{0}{\sqrt{d-1}} = 0 \rightarrow \phi_x - \phi_a \in (-1, 0),
\end{align}
which will be useful in the following section.

Inserting the known form of $\vec{\phi^*}$ (Equation~\ref{eq:phistar}) into $f(\vec{\phi})$ (Equation~\ref{eq:newobj}),
\begin{align}\label{eq:minab}
	f(\vec{\phi}) = f(s_a) = 
		\begin{cases} 
     		s_a \phi_a \log_2 (\phi_a) + (d - s_a)\phi_x \log_2(\phi_x),		 																		& \phi_x > \epsilon \\
     		s_a \phi_a \log_2 (\phi_a) + (d - s_a)\left( \frac{\phi_x^2}{\epsilon 2\ln(2)} + \phi_x\log_2(\epsilon) - \frac{\epsilon}{2\ln(2)}\right), 	&\phi_x \leq \epsilon .
     	\end{cases}
\end{align}
Writing $f(s_a)$ in place of $f(\vec{\phi})$ acknowledges that, as both $\phi_x$ and $\phi_a$ are functions of discrete variable $s_a$, Equation~\ref{eq:minab} is also a single-variable function (of $s_a$). For completeness, we would like to write the regions of the piecewise function in terms of $s_a$, rather than $\phi_x$. To do so, we first study the derivatives of Equation~\ref{eq:ab_a} and Equation~\ref{eq:ab_b}:
\begin{align}
	\frac{\partial \phi_a}{\partial s_a}
		&=	\frac{-\sqrt{\frac{s_a}{d - s_a}\cdot \left(d\mathcal{P}_{MC} - 1\right) }}{2s_a^2}
		= \frac{d\phi_x - 1}{2s_a^2}
		=	\frac{\phi_x - \phi_a}{2s_a}\label{eq:dphia}\\
	\frac{\partial \phi_x}{\partial s_a}
		&=	\frac{-\sqrt{\frac{d - s_a}{s_a} \cdot \left(d\mathcal{P}_{MC} - 1\right)}}{2(d - s_a)^2}
		= \frac{1 - d\phi_a}{2 (d - s_a)^2}
		=	\frac{\phi_x - \phi_a}{2(d - s_a)}.\label{eq:dphib}
\end{align}
As $\phi_x - \phi_a < 0$ and $d-s_a \geq 1$, both derivatives are negative. Consequentially $\phi_a$ and $\phi_x$ are monotonic and thus injective functions of $s_a$, as one might assume. Then it will be simple to convert regions in terms of $\phi_x$ to regions in terms of $s_a$. To find the value of $s_a$ where $\phi_x = \epsilon$ (denoted $s_{a(\phi_x=\epsilon)}$) 
\begin{align}
	&\epsilon
		=	\phi_x = \frac{1 - \sqrt{\frac{s_{a(\phi_x=\epsilon)}}{d - s_{a(\phi_x=\epsilon)}}\cdot \left(d\mathcal{P}_{MC} - 1\right) }   }{d}
	\rightarrow	
	\left(\frac{s_{a(\phi_x=\epsilon)}}{d - s_{a(\phi_x=\epsilon)}}\right) \cdot \left(d\mathcal{P}_{MC} - 1\right)
		=	(1 - d\epsilon)^2\nonumber\\
	\rightarrow&
	s_{a(\phi_x=\epsilon)}
		=
		\frac{(1 - d\epsilon)^2}{\mathcal{P}_{MC} - 2\epsilon + d\epsilon^2} \equiv Q.
\end{align}
Appreciate that $Q$ approaches $1 / \mathcal{P}_{MC}$ from below as $\epsilon \rightarrow 0$. Now the objective function (Equation~\ref{eq:minab}) can be written more formally in terms of $s_a$,
\begin{align}\label{eq:novelminab}
f(s_a) = 
		\begin{cases} 
     		s_a \phi_a \log_2 (\phi_a) + (d - s_a)\phi_x \log_2(\phi_x),		 																		& 1 \leq s_a < Q \\
     		s_a \phi_a \log_2 (\phi_a) + (d - s_a)\left( \frac{\phi_x^2}{\epsilon 2\ln(2)} + \phi_x\log_2(\epsilon) - \frac{\epsilon}{2\ln(2)}\right), 	& d-1 \geq s_a \geq Q.
     	\end{cases}
\end{align}

We have concluded step three of the derivation. We have written the objective function in terms of its $d-1$ potential minima ($s_a = 1, 2, 3, ..., d-1$). In the final step we will determine which of solutions yields the actual constrained minimum.

\subsection{Optimal solution and lower bound}
There are $d-1$ candidate solutions to Equation~\ref{eq:novelminab} represented by $s_a = 1, 2, 3, ..., d-1$. According to the method of Lagrange multipliers, one of these will correspond to the constrained minimum of $f(\vec{\phi})$. That is, if $s_a^*$ is the integer value which minimizes $f(s_a)$, then $f(s_a^*)$ is the constrained minimum of $f(\vec{\phi})$ and the minimizing $\vec{\phi^*}$ is described by Equation~\ref{eq:phistar} with $\phi_a(s_a^*)$ and $\phi_x(s_a^*)$ given by Equation~\ref{eq:ab_a} and Equation~\ref{eq:ab_b}. To determine $s_a^*$, we will treat $s_a$ as a continuous variable and take the derivative of $f(s_a)$ with respect to $s_a$. We will find that $\partial f(s_a) / \partial s_a$ is always positive (on $s_a\in[1, d-1]$), regardless of $d$ or $\mathcal{P}_{MC}$, and therefore in all cases $s_a^* = 1$.

The derivative of Equation~\ref{eq:novelminab} is 
\begin{align}\label{eq:derivative}
	\frac{\partial f(s_a)}{\partial s_a} =
		\begin{cases}			
     		 \log_2(e)\cdot \bigg(
     		 	s_a \frac{\partial \phi_a}{\partial s_a}\left(\ln(\phi_a) + 1 \right)			+ \phi_a\ln(\phi_a)  &\\
     		 	\text{  } + (d - s_a) \frac{\partial \phi_x}{\partial s_a}\left(\ln(\phi_x) + 1  \right)	- \phi_x\ln(\phi_x)
     		 	\bigg),& 1 < s_a < Q \\
     		 \log_2(e)\cdot \bigg(
     		 	s_a \frac{\partial \phi_a}{\partial s_a}\left(\ln(\phi_a) + 1  \right)			+ \phi_a\ln(\phi_a) &\\	
     		 	\text{  }+ (d - s_a)\frac{\partial \phi_x}{\partial s_a}\left(\frac{\phi_x}{\epsilon} + \ln(\epsilon)  \right)	
     		 	-\left(		\frac{\phi_x^2}{2\epsilon} + \phi_x\ln(\epsilon ) - \frac{\epsilon}{2}	\right)
     		 	\bigg), 	&  d-1 \geq s_a \geq Q.
     	\end{cases}
\end{align}
Applying Equation~\ref{eq:dphia} and Equation~\ref{eq:dphib} and simplifying,
\begin{align}\label{eq:derivative2}
	\frac{\partial f(s_a)}{\partial s_a} =
		\begin{cases} 
     		 \log_2(e)\cdot \left(
	     		 \phi_x - \phi_a + \left(\frac{\phi_x + \phi_a}{2}\right) \ln(\phi_a / \phi_x )
    	 		 \right),	& 1 < s_a < Q \\
     		 \frac{\log_2(e)}{2}\cdot \left(
     		 \left(\phi_x - \phi_a\right) + \left(\phi_x + \phi_a\right)\ln(\phi_a / \epsilon) - \left(\frac{\phi_a\phi_x}{\epsilon}\right) + \epsilon
     		 \right),		&  d-1 \geq s_a \geq Q.
     	\end{cases}
\end{align}

First consider the $1 < s_a < Q$ case (i.e. the $\phi_x > \epsilon$ case). Define $x \equiv \phi_a / \phi_x$. Note that as $\phi_a > \phi_x$ in general and as $\phi_x$ is positive in this case (as $\phi_x > \epsilon > 0$) $x$ is strictly greater than one. For $x>1$ the inequality 
\begin{align*}
	\ln(x) > 2\left(\frac{x - 1}{x + 1}\right)
\end{align*}
holds. Consequently, replacing $x$ with $\phi_a / \phi_x$ yields
\begin{align}
		&\ln(\phi_a / \phi_x) > \left(\frac{\phi_x}{\phi_x}\right) \cdot 2 \left(\frac{\phi_a / \phi_x - 1}{\phi_a / \phi_x + 1}\right) = 2\cdot\left(\frac{\phi_a - \phi_x}{\phi_a + \phi_x}\right) \nonumber\\
	\rightarrow&
		 \phi_x - \phi_a + \left(\frac{ \phi_a + \phi_x }{2}\right) \ln(\phi_a / \phi_x) > 0.
\end{align}
Therefore in the $1 < s_a < Q$ case, the derivative of $f(s_a)$ with respect to $s_a$ (Equation~\ref{eq:derivative2}) is strictly positive.

Now examine the second case, where $d-1 \geq s_a \geq Q$ (i.e. $\phi_x \leq \epsilon$). As $\phi_x - \phi_a \in (-1, 0)$ (Equation~\ref{eq:axdifference}), if the quantity
\begin{align}\label{eq:toshow}
-1 + (\phi_x + \phi_a)\ln(\phi_a / \epsilon ) - \left(\frac{\phi_a\phi_x}{\epsilon}\right) + \epsilon > 0
\end{align}
then $\frac{\partial f(s_a)}{\partial s_a}$ is positive for $d-1 \geq s_a \geq Q$ (Equation~\ref{eq:derivative2}). It is sufficient to show
\begin{align}\label{eq:toshow2}
	&(\phi_x + \phi_a)\ln(\phi_a / \epsilon ) - \left(\frac{\phi_a\phi_x}{\epsilon}\right) -1 \nonumber \\
	&= \phi_a\ln(\phi_a / \epsilon )
	+ (-\phi_x ) \left( \frac{\phi_a}{\epsilon} - \ln(\phi_a / \epsilon ) \right)
	-1 \geq 0
\end{align}
for Equation~\ref{eq:toshow} to hold and to prove $\frac{\partial f(s_a)}{\partial s_a}$ is positive.

Again, showing Equation~\ref{eq:toshow2} holds in the region $\phi_x \leq \epsilon$ (i.e. $-\phi_x \geq -\epsilon$) is sufficient to show $f(s_a)$ increases monotonically in that region. Recognize that as $\phi_a > 1/d \gg \epsilon$, 
\begin{align*}
	\frac{\phi_a}{\epsilon} \gg 1
\end{align*}
and consequently the quantity 
\begin{align*}
	\frac{\phi_a}{\epsilon} - \ln\left( \frac{\phi_a}{\epsilon} \right) > 0.
\end{align*}
Make three final notes, (1) that $\ln(\phi_a / \epsilon )$ is positive; (2) that $\phi_a\ln(\phi_a /(e \epsilon ))$ is monotonically increasing with $\phi_a$; and (3) that $-\epsilon\ln(\epsilon)$ is positive (as $\epsilon$ is positive and less than $1$). Then
\begin{align}\label{eq:geqchain}
	&\phi_a\ln(\phi_a / \epsilon ) + (-\phi_x)\left(\frac{\phi_a}{\epsilon} - \ln(\phi_a / \epsilon )\right) - 1
	\geq \phi_a\ln(\phi_a / \epsilon ) - \epsilon\left(\frac{\phi_a}{\epsilon} - \ln(\phi_a / \epsilon )\right) -1 \nonumber \\
	&\geq \phi_a\ln(\phi_a / \epsilon ) - \phi_a +\epsilon \ln(\phi_a) - \epsilon\ln(\epsilon) -1
	\geq \phi_a\left(\ln(\phi_a / \epsilon) - 1\right) +\epsilon\ln(1/d) -1 \nonumber \\
	&= \phi_a\ln\left(\frac{\phi_a}{\epsilon e}\right) +\epsilon\ln(1/d) -1
	\geq \left(\frac{1}{d}\right)\cdot\left( -\ln\left(ed\epsilon\right) - d\epsilon \ln(d) - d   \right) \nonumber \\
	&= \frac{-\ln(e^{1+d} d^{1+d\epsilon}\epsilon)}{d}
\end{align}
If the final quantity in Equation~\ref{eq:geqchain} is positive then Equation~\ref{eq:toshow2} will be positive as well. Equation~\ref{eq:geqchain} is positive whenever
\begin{align*}
	\epsilon < \frac{1}{e^{1+d} d^{1 + d\epsilon }}.
\end{align*}
We need not overanalyze this inequality. As we have already presupposed that $\epsilon \ll 1/d$ we know $d\epsilon \ll 1$. Therefore 
\begin{align}\label{eq:smallep}
	\epsilon < \frac{1}{e^{1+d} d^{1 + \delta}}
\end{align}
(where $\delta$ is small and positive) is (in conjunction with $\epsilon \ll 1/d$, i.e. $\epsilon < \delta/d$) a sufficient condition for Equation~\ref{eq:derivative2} to be positive in the $d-1 \geq s_a \geq Q$ case.

We have shown that -- for $\epsilon \ll 1/d$ (and more explicitly when $\epsilon $ satisfies Equation~\ref{eq:smallep}) -- $f(s_a)$ is a strictly increasing function of $s_a$ in $1 \leq s_a \leq d-1$. As such, $f(s_a)$ achieves its minimum value at $s_a^* = 1$ and $f(s_a = 1)$ is the constrained minimum of $f(\vec{\phi})$ according to the method of Lagrange multipliers.

Thus, applying Equation~\ref{eq:ab_a} (middle equality) and Equation~\ref{eq:ab_b} (right equality) to $f(s_a)$ (Equation~\ref{eq:novelminab}) for $s_a=1$, the constrained minimum is
\begin{align}\label{eq:apFinal}
	f(s_a=1) = f(\vec{\phi^*}) &= \phi_a \log_2(\phi_a) + (1 - \phi_a)\log_2\left(\frac{1 - \phi_a}{d - 1} \right) \\
	&\text{ where } \phi_a = \frac{1 + \sqrt{(d\mathcal{P}_{MC} - 1)(d - 1)}}{d} \nonumber
\end{align}
Appreciate that this solution is physically reasonable for the problem of interest. As $d - 1 > d\mathcal{P}_{MC} - 1$ the quantity $\frac{d\mathcal{P}_{MC} - 1}{d - 1}$ is less than $1$ and
\begin{align}\label{eq:finalx}
	\phi_x = \frac{1 - \sqrt{(d\mathcal{P}_{MC} - 1)/(d - 1)}}{d} > 0
\end{align}
making all elements in $\vec{\phi^*}$ nonnegative, as would be required if these elements were the genuine eigenvalues of the system. More significantly, the minimum occurs in $\mathbb{R}^d_{>0}$, where the actual function of interest, Equation~\ref{eq:task}, is defined.

Most importantly, the minimized value of $f(\vec{\phi})$ is not a function of $\epsilon$, so long as $\epsilon$ is sufficiently small (Equation~\ref{eq:smallep}). Thus we can take $\epsilon$ arbitrarily close to zero and Equation~\ref{eq:newobj} becomes an arbitrarily good approximation of the actual negative entropy function of interest (Equation~\ref{eq:task}). Consequently, the constrained minimum of Equation~\ref{eq:newobj} is the lower bound for Equation~\ref{eq:task}. That is, Equation~\ref{eq:apFinal} is the lower bound of Equation~\ref{eq:task} when $d\geq 3$ and $\mathcal{P}_{MC} \in (1/d, 1)$.

Conveniently, while Equation~\ref{eq:apFinal} is derived for the $d \geq 3$ and $\mathcal{P}_{MC} \in (1/d, 1)$ case, note that it is also the correct form of the solution in the $d=2$ for all $\mathcal{P}_{MC} \in [.5, 1]$ cases and is the correct form of the solution for any value of $d$ in all $\mathcal{P}_{MC} = 1$ and $\mathcal{P}_{MC} = 1/d$ cases (if we as usual let $0\cdot \log_2(0) \equiv 0$). In short, Equation~\ref{eq:apFinal} is the lower bound of Equation~\ref{eq:task} for $d\geq 2$ and $\mathcal{P} \in [1/d, 1]$, i.e. everywhere.

Of course, Equation~\ref{eq:apFinal} is a lower bound determined without consideration of the third constraint, Equation~\ref{eq:majcons}. The method of Lagrange multipliers is designed for equality constraints (e.g. Equation~\ref{eq:eigcons} and Equation~\ref{eq:puritycons}), although it can be generalized to the Kuhn-Tucker problem~\cite{de2000mathematical} to incorporate inequality constraints of the form of Equation~\ref{eq:majcons} to more accurately lower bound Equation~\ref{eq:task}. In cases where purity $\mathcal{P}_{MC}$ is high and the largest eigenvalue $\phi_a$ is consequently large, including Equation~\ref{eq:majcons} may have no or little effect. Nevertheless, it may be useful in some instances and we note it as readily available additional information.

\section{Parity measurement determines purity}\label{appendix:paritypurity}
\beginappendix{C}
For photonic carriers, a beamsplitter followed by a parity measurement will measure purity. 

Let $\hat{a}_{m}^\dagger$ be the creation operator for a photon in mode $m$ in port $a$ of a beamsplitter. Then a 2-photon state with one photon in mode $+m$ and the other photon in mode $-m$ entering port $a$ of a beamsplitter is written
\begin{align}
	\hat{a}_{+m}^\dagger  \hat{a}_{-m}^\dagger \ket{\text{vac}}
\end{align}
and a coherent superposition of such states is
\begin{align}
	\ket{\phi_j^{(a)}} = \sum_m \alpha^{(j)}_m \hat{a}_{+m}^\dagger  \hat{a}_{-m}^\dagger \ket{\text{vac}}
\end{align}
where
\begin{align} 
	\sum_m \left\lvert \alpha^{(j)}_m \right\rvert^2  = 1.
\end{align}
It is understood that $+m$ and $-m$ are orthogonal modes (for integer $m \geq 1$), i.e. $\braket{\text{vac}| \hat{a}_{+m}  \hat{a}_{-m}^\dagger  |\text{vac}} = 0$ and more generally
\begin{align}
	\braket{\text{vac}| \hat{a}_{m}  \hat{a}_{n}^\dagger  |\text{vac}} = \delta_{m,n}.
\end{align}
These modes might be frequency bins, time bins, polarization, OAM modes, or any other photonic degree of freedom. 
Consider a different 2-photon state entering port $b$ of the beamsplitter,
\begin{align}
	\ket{\phi_k^{(b)}} = \sum_m \alpha^{(k)}_m \hat{b}_{+m}^\dagger  \hat{b}_{-m}^\dagger \ket{\text{vac}}.
\end{align}
When written in the same spatial mode ($x$), the inner product of $\ket{\phi_j^{(x)}}$ and $\ket{\phi_k^{(x)}}$ is
\begin{align}\label{eq:innerP}
	\braket{\phi_j^{(x)} | \phi_k^{(x)}} = \braket{\phi_j | \phi_k}
		&= \sum_m \bra{\text{vac}} \alpha^{*(j)}_m \hat{x}_{+m} \hat{x}_{-m} \sum_n \alpha^{(k)}_n \hat{x}_{+n}^\dagger  \hat{x}_{-n}^\dagger \ket{\text{vac}} \\ \nonumber
		&= \sum_m \alpha^{*(j)}_m\alpha^{(k)}_m
\end{align}
where $\alpha^{*(j)}_m$ is the complex conjugate of $\alpha^{(j)}_m$.

Consider $\ket{\phi_j^{(a)}}$ entering port $a$ of the beamsplitter and $\ket{\phi_k^{(b)}}$ entering port $b$ of the beamsplitter. The beamsplitter is 50:50 and performs the transformation
\begin{subequations}
\begin{align}
	\hat{a}^\dagger_m &\xrightarrow[]{\text{BS}} \frac{\hat{c}^\dagger_m + \hat{d}^\dagger_m}{\sqrt{2}} \\
	\hat{b}^\dagger_m &\xrightarrow[]{\text{BS}} \frac{\hat{c}^\dagger_m - \hat{d}^\dagger_m}{\sqrt{2}}.
\end{align}
\end{subequations}
The emerging 4-photon state is
\begin{align}\begin{split} 
\ket{\phi_j^{(a)}}\ket{\phi_k^{(b)}} &\xrightarrow[]{\text{BS}}
	\sum_m\sum_n \frac{\alpha^{(j)}_m\alpha^{(k)}_n}{4} \left(\hat{c}^\dagger_{+m} + \hat{d}^\dagger_{+m}\right)\left(\hat{c}^\dagger_{-m} + \hat{d}^\dagger_{-m}\right)\left(\hat{c}^\dagger_{+n} - \hat{d}^\dagger_{+n}\right)\left(\hat{c}^\dagger_{-n} - \hat{d}^\dagger_{-n}\right)			\ket{\text{vac}} \\
	= &\sum_m \frac{\alpha^{(j)}_m\alpha^{(k)}_m}{4} \left(\hat{c}^{\dagger 2}_{+m} - \hat{d}^{\dagger 2}_{+m}\right)\left(\hat{c}^{\dagger 2}_{-m} - \hat{d}^{\dagger 2}_{-m}\right)			\ket{\text{vac}} \\
	&+ \sum_m \sum_{n<m} \frac{\alpha^{(j)}_m\alpha^{(k)}_n + \alpha^{(k)}_m\alpha^{(j)}_n}{4} \bigg(
				\hat{c}^\dagger_{+m}\hat{c}^\dagger_{+n}\hat{c}^\dagger_{-m}\hat{c}^\dagger_{-n}
				- \hat{c}^\dagger_{+m}\hat{c}^\dagger_{+n}\hat{d}^\dagger_{-m}\hat{d}^\dagger_{-n} \\
	&\text{   } -\hat{d}^\dagger_{+m}\hat{d}^\dagger_{+n}\hat{c}^\dagger_{-m}\hat{c}^\dagger_{-n} 
				+ \hat{d}^\dagger_{+m}\hat{d}^\dagger_{+n}\hat{d}^\dagger_{-m}\hat{d}^\dagger_{-n}
				+\hat{c}^\dagger_{+m}\hat{d}^\dagger_{+n}\hat{c}^\dagger_{-m}\hat{d}^\dagger_{-n}
				- \hat{c}^\dagger_{+m}\hat{d}^\dagger_{+n}\hat{d}^\dagger_{-m}\hat{c}^\dagger_{-n} \\
	&\text{   }	-\hat{d}^\dagger_{+m}\hat{c}^\dagger_{+n}\hat{c}^\dagger_{-m}\hat{d}^\dagger_{-n} 
				+ \hat{d}^\dagger_{+m}\hat{c}^\dagger_{+n}\hat{d}^\dagger_{-m}\hat{c}^\dagger_{-n}
				\bigg) \ket{\text{vac}} \\
	&+ \sum_m \sum_{n<m} \frac{\alpha^{(j)}_m\alpha^{(k)}_n - \alpha^{(k)}_m\alpha^{(j)}_n}{4} \bigg(
				-\hat{c}^\dagger_{+m}\hat{c}^\dagger_{+n}\hat{c}^\dagger_{-m}\hat{d}^\dagger_{-n} 
				+ \hat{c}^\dagger_{+m}\hat{c}^\dagger_{+n}\hat{d}^\dagger_{-m}\hat{c}^\dagger_{-n}  \\
	&\text{   } + \hat{d}^\dagger_{+m}\hat{d}^\dagger_{+n}\hat{c}^\dagger_{-m}\hat{d}^\dagger_{-n} 
				- \hat{d}^\dagger_{+m}\hat{d}^\dagger_{+n}\hat{d}^\dagger_{-m}\hat{c}^\dagger_{-n}
				-\hat{c}^\dagger_{+m}\hat{d}^\dagger_{+n}\hat{c}^\dagger_{-m}\hat{c}^\dagger_{-n} 
				+ \hat{c}^\dagger_{+m}\hat{d}^\dagger_{+n}\hat{d}^\dagger_{-m}\hat{d}^\dagger_{-n} \\
	&\text{   }	+\hat{d}^\dagger_{+m}\hat{c}^\dagger_{+n}\hat{c}^\dagger_{-m}\hat{c}^\dagger_{-n}
				- \hat{d}^\dagger_{+m}\hat{c}^\dagger_{+n}\hat{d}^\dagger_{-m}\hat{d}^\dagger_{-n}
				\bigg) \ket{\text{vac}}.
\end{split}\end{align}
The first two quantities represent even parity events: events where either two photons are measured in port $c$ and 2 in port $d$ or all four photons are measured in port $c$ ($d$) and zero are measured in port $d$ ($c$). The final summation contains only odd parity events: events where three photons are measured from port $c$ ($d$) and one from port $d$ ($c$). Assigning even parity events a value of $+1$ and odd parity events the value $-1$ the expectation value of a parity measurement is
\begin{align}
\left\langle \text{parity}_{\ket{\phi_j}, \ket{\phi_k}}\right\rangle 
	&=\sum_m (+1)\cdot 4\left\lvert\frac{\sqrt{2}^2 \alpha^{(j)}_m\alpha^{(k)}_m}{4}\right\rvert^2 \nonumber \\
	&+ \sum_m\sum_{n<m} (+1)\cdot 8\left\lvert\frac{\alpha^{(j)}_m\alpha^{(k)}_n + \alpha^{(k)}_m\alpha^{(j)}_n}{4}\right\rvert^2
	+ (-1)\cdot 8\left\lvert\frac{\alpha^{(j)}_m\alpha^{(k)}_n - \alpha^{(k)}_m\alpha^{(j)}_n}{4}\right\rvert^2 \nonumber\\
	&= 	\sum_m \left\lvert\alpha^{(j)}_m\alpha^{(k)}_m\right\rvert^2
	+ \sum_m\sum_{n<m} \frac{\left\lvert\alpha^{(j)}_m\alpha^{(k)}_n + \alpha^{(k)}_m\alpha^{(j)}_n\right\rvert^2
	- \left\lvert\alpha^{(j)}_m\alpha^{(k)}_n - \alpha^{(k)}_m\alpha^{(j)}_n\right\rvert^2}{2} \nonumber\\
	&= 	\sum_m \left\lvert\alpha^{(j)}_m\alpha^{(k)}_m\right\rvert^2
	+ \sum_m\sum_{n<m} \frac{2\alpha^{(j)}_m\alpha^{(k)}_n\alpha^{*(k)}_m\alpha^{*(j)}_n + 2\alpha^{*(j)}_m\alpha^{*(k)}_n\alpha^{(k)}_m\alpha^{(j)}_n}{2} \nonumber\\
	&= 	\sum_m \left\lvert\alpha^{(j)}_m\alpha^{(k)}_m\right\rvert^2
	+ \sum_m\sum_{n\neq m} \alpha^{(j)}_m\alpha^{(k)}_n\alpha^{*(k)}_m\alpha^{*(j)}_n \nonumber\\
	&= 	\sum_m \left[ \alpha^{(j)}_m\alpha^{(k)}_m\alpha^{*(j)}_m\alpha^{*(k)}_m + \sum_{n\neq m} \alpha^{(j)}_m\alpha^{(k)}_n\alpha^{*(k)}_m\alpha^{*(j)}_n \right] \nonumber \\
	&= 	\sum_m \left[\alpha^{(j)}_m\alpha^{*(k)}_m \left( \alpha^{(k)}_m\alpha^{*(j)}_m + \sum_{n\neq m} \alpha^{(k)}_n\alpha^{*(j)}_n \right) \right] \nonumber \\
	&= 	\left(\sum_m \alpha^{(j)}_m\alpha^{*(k)}_m \right) \left(\sum_{n} \alpha^{(k)}_n\alpha^{*(j)}_n\right) \nonumber \\ 
	&\text{applying Equation~\ref{eq:innerP},} \nonumber \\
	&= \braket{\phi_k | \phi_j} \cdot \braket{\phi_j | \phi_k} \nonumber \\
	&= \left\lvert \braket{\phi_k | \phi_j} \right\rvert^2.
\end{align}
Showing that when $\braket{\phi_k | \phi_j} = 1$ (i.e. $\ket{\phi_k}$ and $\ket{\phi_j}$ have identical mode structure) the expectation value of the parity measurement is $1$: only even-parity events will occur. When $\braket{\phi_k | \phi_j} = 0$ (i.e. $\ket{\phi_k}$ and $\ket{\phi_j}$ are orthogonal in mode structure) the expectation value of the parity measurement is $0$: an equal number of even- and odd-parity events will occur. This concludes analysis for two pure 2-photon states entering either port of a beamsplitter.

With this result in mind, consider the case of an arbitrary (not necessarily pure) 2-photon state $\rho$ with eigendecomposition
\begin{align*}
	&\rho = \sum_k \phi_k \ket{\phi_k}\bra{\phi_k} \tag{\ref{eq:eigendecomp}}
\end{align*}
where
\begin{align*}
	&\braket{\phi_j | \phi_k} = \delta_{j,k} \tag{\ref{eq:eigenterm2}}
\end{align*}
and $\phi_k$ are the nonnegative eigenvalues which sum to one. Consider $\rho$ entering port $a$ of a beamsplitter and an identically prepared $\rho$ simultaneously entering port $b$ of the beamsplitter. This can be treated as though $\ket{\phi_j^{(a)}}$ enters port $a$ with probability $\phi_j$ and $\ket{\phi_k^{(b)}}$ enters port $b$ with probability $\phi_k$. The expectation value of the parity measurement is then (applying Equation~\ref{eq:puritycons})
\begin{align}
	\left\langle \text{parity}_{\rho, \rho}\right\rangle
		&= \sum_j\sum_k \phi_j\phi_k \left\langle \text{parity}_{\ket{\phi_j}, \ket{\phi_k}}\right\rangle  = \sum_j\sum_k \phi_j\phi_k \left\lvert \braket{\phi_k | \phi_j} \right\rvert^2 \nonumber \\
		&= \sum_j\sum_k \phi_j\phi_k \delta_{j,k} = \sum_k \phi_k^2 \nonumber \\
		&= \mathcal{P}(\rho).
\end{align}

We have shown that the purity of $\rho$ can be determined by a 4-photon parity measurement when two identically prepared $\rho$ are simultaneously available. Preparation of two $\rho$ has made the nonlinear measurement of purity ($\tr(\rho^2 )$) a linear measurement and rendered it experimentally accessible.

We note in the most general case where state $\rho_1$ enters port $a$ and state $\rho_2$ enters port $b$ of the beamsplitter where
\begin{subequations}
\begin{align}
	\rho_1 &= \sum_k \phi_k \ket{\phi_k}\bra{\phi_k}  \\
	\rho_2 &= \sum_j \psi_j \ket{\psi_j}\bra{\psi_j} ,
\end{align}
\end{subequations}
the expected parity measurement is
\begin{align}
	\left\langle \text{parity}_{\rho_2, \rho_1}\right\rangle = \sum_j\sum_k \psi_j\phi_k \left\lvert\braket{\phi_k  | \psi_j}\right\rvert^2
\end{align}
which is less than or equal to $\max\left\{\mathcal{P}(\rho_1), \mathcal{P}(\rho_2)\right\}$. Thus in an imperfect experiment the expectation value of the purity will always lower bound the purity of the most pure prepared state.

\end{document}